# Aggregation Model and Market Mechanism for Virtual Power Plant Participation in Inertia and Primary Frequency Response

Changsen Feng, *Member, IEEE*, Zhongliang Huang, Jun Lin, Licheng Wang, Youbing Zhang, *Senior Member, IEEE*, and Fushuan Wen, *Fellow, IEEE*

***Abstract*-- The declining provision of inertia by synchronous generators in modern power systems necessitates aggregating distributed energy resources (DERs) into virtual power plants (VPPs) to unlock their potential in delivering inertia and primary frequency response (IPFR) through ancillary service markets. To facilitate DER participation in the IPFR market, this paper proposes an aggregation model and market mechanism for VPPs participating in IPFR. First, an energy-reserve-IPFR market framework is developed, in which a VPP acts as an intermediary to coordinate heterogeneous DERs. Second, by taking into account the delay associated with inertial response, an optimization-based VPP aggregation method is introduced to encapsulate the IPFR process involving a variety of DERs. Third, an energy-reserve-IPFR market mechanism with VPP participation is introduced, aiming to minimize social costs, where stochastic deviations of renewable energy generation are explicitly modeled through chance-constrained reformulations, ensuring that the cleared energy, reserve, and IPFR schedules remain secure against forecast errors. Case studies on IEEE 30-bus and IEEE 118-bus systems show that the nadir and quasi-steady-state frequencies are reproduced by the VPP aggregation model with a mean absolute percentage error ≤ 0.03%, and the proposed market mechanism with VPP participation reduces the total system cost by approximately 40% and increases the net profit by about 30%.**

## NOMENCLATURE

### *A. Acronyms*

| | |
|---|---|
| ES | Energy storage |
| EV | Electric vehicle |
| FL | Flexible load |
| GFM | Grid-forming |
| IBR | Inverter-based resource |
| IPFR | Inertia and primary frequency response |
| QSS | Quasi-steady-state |
| PFR | Primary frequency response |
| PWL | Piecewise linearization |
| REG | Renewable energy generator |
| RoCoF | Rate-of-change-of-frequency |
| SG | Synchronous generator |
| VPP | Virtual power plant |

### *B. Index*

| | |
|---|---|
| $\upsilon$, $\omega$ | Index of VPP aggregated order |
| $hp$ | Index of PWL planes |
| $i, j, k, l$ | Index of SGs, GFM inverter-based REGs, GFM inverter-based ES units, VPPs |
| $n, m$ | Index of buses in the system |
| $v$ | Index of internal units within VPP |

### *C. Sets*

| | |
|---|---|
| $\Omega_{G/ESS/REG/VPP}$ | Set of SGs/ES units/REGs/VPPs |
| $\mathcal{T}$ | Set of hourly periods of the SGs offline time span before start-up |
| $N_{G/FM}$ | Set of SGs/GFM IBRs within the VPP |
| $N_{hp}$ | Set of PWL planes |

### *D. Parameters*

| | |
|---|---|
| $\alpha_{i/j/k,t}^{G/REG/ESS}$ | Unit power generation costs of SG $i$/REG $j$ /ES $k$ at time $t$ [\$/MWh] |
| $\alpha_{l,t}^{VPPG}$, $\alpha_{l,t}^{VPPR}$ | Unit power generation costs of SGs/ converter-interfaced DERs within the VPP $l$ at time $t$ [\$/MWh] |
| $\beta_{j/k/l,t}^{REG/ESS/VPPR}$ | Unit virtual inertia costs of REG $j$/ES $k$ /VPP $l$ at time $t$ [\$/(MW·s/Hz)] |
| $\chi_{j/k/l,t}^{REG/ESS/VPP}$ | Unit droop factor costs of REG $j$/ ES $k$ /VPP $l$ [\$/(MW/Hz)] |
| $\rho_{i/j/k/l,t}^{G/REG/ESS/VPP,EN}$ | Price of SG $i$/REG $j$/ES $k$/VPP $l$ providing unit power at time $t$ [\$/MWh] |
| $\rho_{i/j/k/l,t}^{G/REG/ESS/VPP,IN}$ | Price of SG $i$/REG $j$/ES $k$/VPP $l$ providing unit inertia at time $t$ [\$/(MW·s/Hz)] |
| $\rho_{i/j/k/l,t}^{G/REG/ESS/VPP,DF}$ | Price of SG $i$/REG $j$/ES $k$/VPP $l$ providing unit droop factor at time $t$ [\$/(MW/Hz)] |
| $\boldsymbol{\varepsilon}_{j/sys,t}$ | Vector of random forecast error of REG $j$/ system power output [MW] |
| $\mu_{j/sys,t}$, $\sigma_{j/sys,t}^{2}$ | Mean and variance of random forecast error $\boldsymbol{\varepsilon}_{j/sys,t}$ [MW] |
| $\mu_k^{c/d}$ | Charging/discharging efficiency of ES $k$ |
| $\kappa_{hp}$ | Parameters of PWL planes |

This work is supported by National Natural Science Foundation of China (No. 52107129 and U22B20116).

C. Feng, Z. Huang, J. Lin, L. Wang and Y. Zhang are with College of Information Engineering, Zhejiang University of Technology, Hangzhou 310023, China (e-mail: fcs@zjut.edu.cn, 201906080109@zjut.edu.cn, 211122030080@zjut.edu.cn, wanglicheng@zjut.edu.cn, youbingzhang@zjut.edu.cn).

F. Wen is with the School of Electrical Engineering, Zhejiang University, Hangzhou 310027, China, and F. Wen is also with the Hainan Institute, Zhejiang University, Sanya 572000, China (e-mail: fushuan.wen@gmail.com).

$\lambda_v$ Normalized gain of the feedback loop in the frequency response model of unit $v$

$\lambda_{nadir}$ , $\lambda_{qss}$ Penalty coefficients for the nadir frequency and QSS frequency

$\tau_1$ , $\tau_2$ Time delay in IPFR for GFM IBR, PFR for SG [s]

$\Phi_A$ Constant coefficient of operational power for the FL

$C(s)$ Thermodynamic equivalent constant of the FL compressor

$C_i^{SU-N/M}$ Fixed/Linear cost for the start-up of SG $i$ [$]

$C_i^{SD}$ Fixed cost for the shut-down of SG $i$ [$]

$\Delta D$ System disturbance [MW]

$D_G$ Damping of the small SG within VPP

$\Delta f_{\text{nadir}-\max}$ Maximum acceptable nadir frequency deviation [Hz]

$\Delta f_{\text{qss}}^{\text{ref}}$ Maximum acceptable QSS frequency deviation [Hz]

$F_h$ Power ratio of the high-pressure cylinder

$G_{VPP}(s)$ Aggregated transfer function of PFR for VPP

$H_{G/FM}$ (Virtual) Inertia of SGs/GFM IBRs [MW·s/Hz]

$H_{VPPG/VPPR}$ (Virtual) Inertia provided by small SGs /GFM IBRs within VPP [MW·s/Hz]

$k_{G/FM/EV/FL/VPP}$ Droop factor of small SGs/GFM IBRs/EVs/FLs/VPPs [MW/Hz]

$k_\upsilon$ , $h_\omega$ $\upsilon$, $\omega$ order coefficients of the numerator and denominator in the polynomial

$K_{m,v}$ Capacity proportion of unit $v$ within VPP

$p_{i/l}^{G/VPP}$ Probability of power constraint violations for SG $i$/ VPP $l$

$p_k^{c/d}$ Probability of charging/ discharging power constraint violations for ES $k$

$P_{i,t,\min}^{G,EN}$ , $P_{i,t,\max}^{G,EN}$ Minimum and maximum generated power of SG $i$ in period $t$ [MW]

$P_{i/l}^{Ramp}$ Maximum ramping rate of SG $i$/small SG within VPP $l$

$P_{j,\max}^{REG,EN/IN/DF}$ Maximum capacity for providing energy/ virtual inertia/droop factor of REG $j$ in period $t$ [MW]

$P_{j,\min}^{REG,EN/IN/DF}$ Minimum capacity for providing energy/ virtual inertia/droop factor of REG $j$ in period $t$ [MW]

$P_{k,\max}^{ESS,c/d}$ Maximum charging/ discharging power of ES $k$ in period $t$ [MW]

$P_{k,\max}^{ESS,IN/DF}$ Maximum reserve capacity for providing virtual inertia/droop factor of ES $k$ in period $t$ [MW]

$P_{k,\min}^{ESS,IN/DF}$ Minimum reserve capacity for providing virtual inertia/droop factor of ES $k$ in period $t$ [MW]

$P_{l,t,\min}^{VPPR,IN/VPP,DF}$ Minimum reserve capacity for providing virtual inertia/droop factor of VPP $l$ [MW]

$P_{l,t,\max}^{VPPR,IN/VPP,DF}$ Maximum reserve capacity for providing virtual inertia/droop factor of VPP $l$ [MW]

$RoCoF_{\max}$ Maximum acceptable RoCoF [Hz/s]

$S_{k,\min/\max}^{ESS}$ Lower and upper limit of energy level of ES $k$ [MWh]

$\Delta t^{IPFR}$ Number of seconds of IPFR process [s]

$T_G$ , $T_R$ , $T_C$ Time constants of the governor, reheater, and steam volume of small SGs [s]

$T_{FM/EV/FL/A}$ Time constant of the GFM inverter response/EV charging/FL compressor/FL thermodynamic [s]

## E. Variables

$\theta_{n,t}$ Angle of node $n$ at time $t$ [rad]

$B_{n,m}$ Susceptance between nodes $n$ and $m$ [S]

$D_{n,t}$ Power consumed of node $n$ at time $t$ [MW]

$\Delta f(t)$ , $\Delta\tilde{f}(t)$ Function of system frequency deviation using the multi-machine model and aggregation model [Hz]

$h_{i,t}^{SU}$ Number of hours that SG $i$ has been offline in start-up time $t$ [h]

$H^{TO/GV}$ Total inertia/non-delayed inertia of the system [MW·s/Hz]

$H_{i/l,t}^{G/VPPG}$ Non-delayed inertia provided by SG $i$/VPP $l$ at time $t$ [MW·s/Hz]

$H_{j/k/l,t}^{REG/ESS/VPPR}$ Virtual inertia provided by REG $j$/ES $k$ /VPP $l$ at time $t$ [MW·s/Hz]

$k_{i/j/k/l,t}^{G/REG/ESS/VPP}$ Droop factor provided by SG $i$/REG $j$/ES $k$ /VPP $l$ at time $t$ [MW/Hz]

$\Delta P_{i/j/k/l,t}^{G/REG/ESS/VPP}$ Power increment of SG $i$/REG $j$/ES $k$/VPP $l$ at time $t$ in the energy market [MW]

$P_{i/j,t}^{G/REG,EN}$ Power provided by SG $i$/REG $j$ at time $t$ in the energy market [MW]

$\boldsymbol{P}_{i/j,t}^{G/REG,EN}$ Random power provided by SG $i$/REG $j$ at time $t$ in the energy market [MW]

$P_{k,t}^{ESS,c/d}$ Power of ES $k$ charging/discharging at time $t$ in the energy market [MW]

$\boldsymbol{P}_{k,t}^{ESS,c/d}$ Random power of ES $k$ charging/ discharging at time $t$ in the energy market [MW]

$P_{l,t}^{VPPG/VPPR,EN}$ Power provided by SGs/converter-interfaced DERs within the VPP $l$ at time $t$ in the energy market [MW]

$\boldsymbol{P}_{l,t}^{VPPG/VPPR,EN}$ Random power provided by SGs/ converter-interfaced DERs within the VPP $l$ at time $t$ in the energy market [MW]

$S_{k,t}^{ESS}$ Energy level of ES $k$ at time $t$ [MWh]

$t_{nadir}$ , $t_{nadir}^*$ Frequency nadir times derived by multi-machine model and aggregation model [s]

$u_{i/l,t}$ Balancing reserve factor of SG $i$/ VPP $l$ power output

$u_{k,t}^{c/d}$ Balancing reserve factor of ES $k$ charging/discharging power output

$x_{i/l,t}$ Status of SG $i$/small SGs within the VPP $l$ indicating start-up/shut-down states at time $t$

$y_{i,t}^{SU/SD}$ Status of SG $i$ indicating start-up/shut-down actions at time $t$

$z_k$ Status of ES $k$ indicating charging states at time $t$

## I. Introduction

In modern power systems, the high penetration of renewable energy resources has led to a significant reduction in system inertia, increasing the risk of a rapid frequency drop following disturbances, which poses critical challenges to frequency stability [1], [2]. For instance, during the South Australian grid blackout in September 2016, the loss of 456 MW power generation within 8 seconds caused the frequency to drop to 47 Hz and the maximum rate-of-change-of-frequency (RoCoF) reached 6.25 Hz/s, far beyond the frequency security boundary of the power system [3]. Similar events have been observed in other countries, such as the United States [4] and the United Kingdom [5].

To address these challenges, it is essential to establish an ancillary market to support system frequency during disturbances [6]. The frequency regulation services typically consist of inertia and primary frequency response (IPFR), which can provide the immediate response to a disturbance, decreasing the frequency decline rate to avoid generator and load outages [7]. Existing ancillary markets primarily focus on either inertial response or primary frequency response (PFR). For example, EirGrid and National Grid Electricity System Operator (NGESO) currently purchase inertia as an ancillary service product [8]. Meanwhile, the Electric Reliability Council of Texas (ERCOT) operates a PFR market [9] and the Australian Energy Market Operator (AEMO) is integrating PFR resources into the contingency frequency services and promoting inertia participation in ancillary markets [10].

In the academic field, researchers have explored IPFR market mechanisms extensively. For instance, a virtual inertia market for inverter-based equipment is presented in [11]. To address the stochastic nature of renewable energy generator (REG) outputs, an inertia pricing scheme in a stochastic market is introduced in [12], which considers non-synchronous resources providing virtual inertia. Similarly, the necessity of PFR markets is discussed in [13], and a pricing mechanism for PFR, accounting for post-disturbance transmission constraints, is developed in [14] to ensure the deliverability of procured reserves. To ensure fair settlement of PFR resources, a market mechanism featuring marginal prices reflecting contributions to different frequency security criteria is proposed in [15]. However, these studies focus either on inertia or PFR, while in reality, inertia and PFR processes are temporally correlated. To address this issue, an IPFR market mechanism is proposed in [16] based on the response rates of synchronous generators (SGs) to quantify the effectiveness of frequency response. Furthermore, IPFR markets are expanded by incorporating a variety of flexible resources, such as variable REGs [17], energy storage (ES) [18], converter-interfaced generators [19], and inverter-based resources [20]. In addition, factors such as the joint energy-IPFR market mechanism [17], delay in inertial response [18], and recovery effects [20] are considered to enhance the performance of IPFR market mechanisms.

However, distributed energy resources (DERs) face entry barriers to participating in the IPFR market due to their limited individual capacities [21]. Virtual power plants (VPPs) emerge as important enablers by aggregating numerous DERs to achieve grid-scale frequency response capabilities in the IPFR market [22], leveraging their collective flexibility and controllability for frequency support [23]. Nevertheless, VPP participation in IPFR markets is hindered by two technical issues: 1) the technical heterogeneity of DERs, which leads to diverse IPFR characteristics, including varying inertial response delays; and 2) the diverse dynamics of DERs bring a high-dimensional parameter space when aggregated [24]. Thus, how to develop an equivalent aggregation model that precisely captures these inertial response delays with a limited number of parameters is a crucial challenge.

Various aggregation models for VPP frequency responses have been studied in recent studies, covering different DER types and modeling methods. An analytical approach to consolidate the frequency response model of a multi-machine system into a single-machine model is presented in [25]. An equivalent aggregation model is derived in [26] based on polynomial fitting of PFR characteristic curves of generator units. To better capture DER-level dynamics, control-loop equivalence is adopted for aggregation in [27], and stochastic tripping is introduced in [28]. However, these methods handle heterogeneity through fitting or averaging, which often fails to accurately represent the criteria of frequency security in power systems. To address this limitation, the authors of [29] consider IPFR delays among diverse DERs and aggregate parameters of frequency response using an optimization framework. Additionally, a stepwise linear programming algorithm is employed in [30] to enhance the accuracy of frequency response aggregation models, considering feedback control branches of DERs. Nonetheless, these approaches fail to capture the heterogeneity among a variety of DERs, particularly their varied inertial response delays. Therefore, a more accurate equivalent aggregation model for VPP frequency responses is still needed.

To fill the above research gap, a joint energy-reserve-IPFR market mechanism is proposed, ensuring secure and delay-aware VPP participation in frequency support while considering REGs' generation uncertainty. The key features of the market mechanism in the existing literature and our paper are compared in Table I. Besides, an optimization-based aggregation model is used to accurately capture the IPFR process of VPPs, accounting for the heterogeneity among diverse DERs. The key features of aggregation methods for VPP in the existing literature and our paper are compared in Table II.

TABLE I
Comparison of the Considered Factors in the IPFR Market Mechanism in the Existing Literature and This Paper

| Refs. | Energy Market clearing | Correlation of Inertia and PFR | REG Uncertainty | Inertial Response Delay | VPP Participation |
|---|---|---|---|---|---|
| [11] | × | × | × | × | × |
| [12] | ✓ | × | ✓ | × | × |
| [13], [14], [15] | ✓ | × | × | × | × |
| [16] | ✓ | ✓ | × | × | × |
| [17] | ✓ | ✓ | ✓ | × | × |
| [18] | ✓ | ✓ | × | ✓ | × |
| [19] | ✓ | ✓ | ✓ | × | × |
| [20] | ✓ | ✓ | ✓ | ✓ | × |
| This paper | ✓ | ✓ | ✓ | ✓ | ✓ |

TABLE II
COMPARISON OF THE CONSIDERED FACTORS IN THE VPP AGGREGATION MODEL IN THE EXISTING LITERATURE AND THIS PAPER

| Refs. | Inertia Aggregation | PFR Aggregation | Optimization Method | Heterogeneous Response | Inertial Response Delay |
|---|---|---|---|---|---|
| [25], [26] | ✓ | ✓ | × | × | × |
| [27] | ✓ | ✓ | × | ✓ | × |
| [28] | × | ✓ | × | × | × |
| [29] | × | ✓ | ✓ | × | ✓ |
| [30] | ✓ | ✓ | ✓ | ✓ | × |
| This paper | ✓ | ✓ | ✓ | ✓ | ✓ |

The contributions of this paper are summarized as follows:

1) An optimization-based aggregation model is proposed to characterize the IPFR process of VPPs. The model classifies the aggregated inertia into non-delayed and delayed components and consolidates the heterogeneous PFR dynamics of diverse DERs into a unified representation. The model's parameters are determined via weighted averaging and stochastic gradient descent (SGD), which yields a compact, low-order representation of the aggregated VPP's IPFR.

2) A dynamic frequency response framework is developed that incorporates the distinct inertial response characteristics of SGs, grid-forming (GFM) inverter-based resources (IBRs), and the proposed VPP model. The framework models the multi-stage system frequency dynamics by distinguishing between instantaneous inertial response and delayed PFR. Frequency-security constraints are derived and integrated into the market-clearing process, ensuring that procured IPFR services account for their specific response timing.

3) A joint energy-reserve-IPFR market mechanism is designed considering REGs' generation uncertainty, enabling VPPs to participate as aggregated entities by bidding equivalent inertia and droop factors derived from the proposed aggregation model. Energy, reserve, and IPFR services are co-optimized, with chance-constrained reformulations employed to manage REGs' generation uncertainty. Case studies on IEEE 30-bus and 118-bus systems are presented, validating the secure and economic participation of VPPs in low-inertia systems.

## II. IPFR ANCILLARY SERVICES FRAMEWORK

In this section, an IPFR market framework is presented accommodating device-specific delays, thereby enabling effective participation of VPPs in the provision of IPFR services.

### A. Ancillary Service for IPFR

When a power system experiences a disturbance, it is essential to implement effective measures to restore frequency to a steady state. To characterize the system's frequency-security margin following a disturbance, three key metrics are considered [31]: the maximum acceptable RoCoF $RoCoF_{\max}$, the maximum acceptable nadir frequency deviation $\Delta f_{\text{nadir-max}}$, and the maximum acceptable quasi-steady-state (QSS) frequency deviation $\Delta f_{\text{qss}}^{\text{ref}}$. To ensure system security, these metrics are commonly set to [15]: $RoCoF_{\max}$ = 0.125Hz/s; $\Delta f_{\text{nadir-max}}$ = 0.5Hz; $\Delta f_{\text{qss}}^{\text{ref}}$ = 0.25Hz.

As the issue of low inertia in modern power systems becomes increasingly evident, the need for frequency ancillary service rises, particularly in the very beginning stages of disturbances, which makes IPFR even more crucial.

The IPFR process in traditional power systems consists of an inertial response followed by a PFR process. The inertial response process is governed by the rotor swing equation of the SGs:

$$2H_i^G \cdot \frac{d\Delta f(t)}{dt} = \Delta P_i^G - \Delta D, i \in \Omega_G \tag{1}$$

Then, the governor ramps the turbine's mechanical power, and the generator converts this incremental mechanical energy into electrical energy that supports the PFR action:

$$T_i^G \cdot \frac{d\Delta P_{i,t}^G}{dt} + \Delta P_{i,t}^G = -k_i^G \Delta f(t), i \in \Omega_G, t \geq \tau_2 \tag{2}$$

Moreover, newly developed technologies such as VPPs and GFM IBRs are capable of participating in IPFR by providing (virtual) inertia and droop factors. The timeline of IPFR across different resources is illustrated in Fig. 1. To avoid unnecessary triggering during minor frequency fluctuations, a short response delay $\tau_1$ is usually adopted in the control loops of GFM IBRs. Besides, SGs exhibit a longer PFR delay $\tau_2$ due to the governor dynamics. Notably, VPPs can aggregate DERs with complementary characteristics—combining the instantaneous inertial response from small SGs with the rapidly dispatchable IPFR provided by converter-interfaced DERs—thereby facilitating effective participation across the entire IPFR timeline.

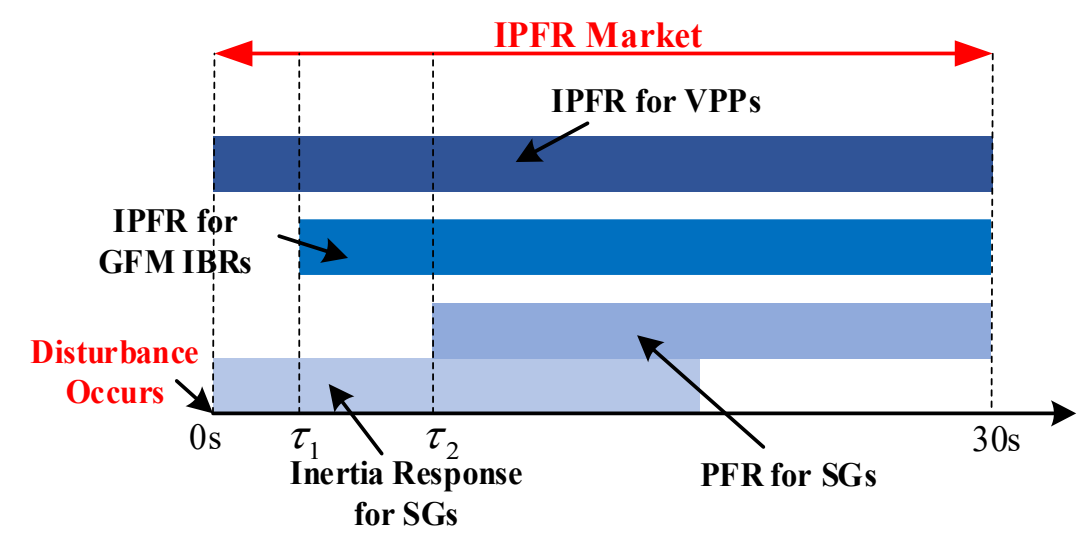

Fig. 1. The timeline of IPFR for different types of resources.

### B. IPFR Market Framework with VPP Participation

The provision of IPFR is inherently coupled with real-time energy production in SGs, necessitating a joint energy-IPFR market. Meanwhile, the power generation uncertainty from variable REGs requires an explicit reserve market to ensure real-time power balance of the system. Therefore, a joint energy-reserve-IPFR market is established as shown in Fig. 2. VPPs are key participants in this market, acting as a single bidding entity by aggregating small SGs, REGs, ES units, electric vehicles (EVs), flexible loads (FLs), and other DERs to leverage their combined characteristics.

In the joint energy-reserve-IPFR market, participants consist of the system operator (SO), service providers, and response resources. Service providers submit their bids to the SO, which consist of offering curves for the energy and reserve markets, as well as bids for inertia and droop factors in the IPFR market. The SO then executes market-clearing while accounting for frequency-security constraints. Following the market-clearing results, service providers dispatch corresponding control signals to the response resources.

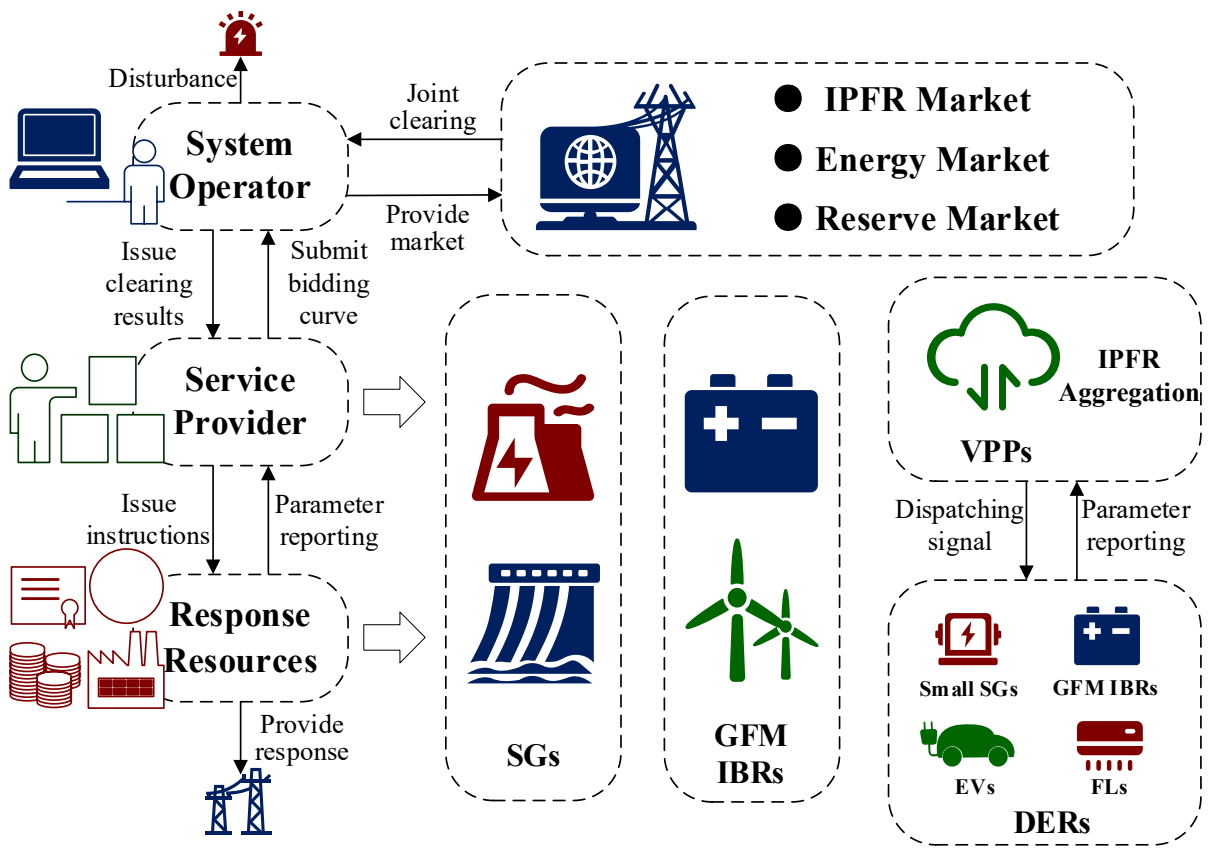

Fig. 2. The joint energy-reserve-IPFR market framework with VPP participation.

Among the market participants, SGs and GFM IBRs serve as both service providers and response resources, while VPPs participate exclusively as service providers. Specifically, the IPFR capabilities of the DERs are aggregated by the VPP to derive equivalent inertia and droop factors before market opening. Then, given the instantaneous requirements for inertial response, VPPs submit their bids for non-delayed/ delayed inertia and aggregated droop factors. Following market clearance, VPPs the procured products to the specific technical parameters of individual DERs to determine the optimal resource allocation for actual service delivery.

## III. The Aggregation Model of VPP

In this section, an aggregation model is proposed considering the frequency response characteristics of heterogeneous DERs within a VPP.

### *A. Frequency Response Models of Multiple DERs*

A diverse portfolio of adjustable DERs—small SGs, GFM IBRs, EVs, FLs, and any other resources capable of delivering inertial response or PFR—is aggregated into the VPP.

The inertial response of small SGs is described in (1) and the PFR of small SGs is derived from the governor of generators [32], as illustrated in Fig. 3(a).

GFM IBRs achieves IPFR through inverter control. A common strategy is the virtual SG (VSG) approach, which emulates SG rotor dynamics to provide virtual inertia and droop factors [33] with a delay $\tau_1$, avoiding unnecessary triggering. By incorporating a delay module, the IPFR model controlled by VSG is shown in Fig. 3(b).

Usually, EV clusters are pooled into a single demand-side resource and connected to charging stations to provide potential system PFR [34]. The PFR model of EV clusters is shown in Fig. 3(c).

Continuously adjustable FLs, which use converters and inverters (e.g., inverter-based air conditioners), can be controlled by adjusting the motors' or compressors' operating frequency to regulate power consumption continuously [35]. This capability enables them to provide PFR, as shown in the PFR model of adjustable FLs in Fig. 3(d).

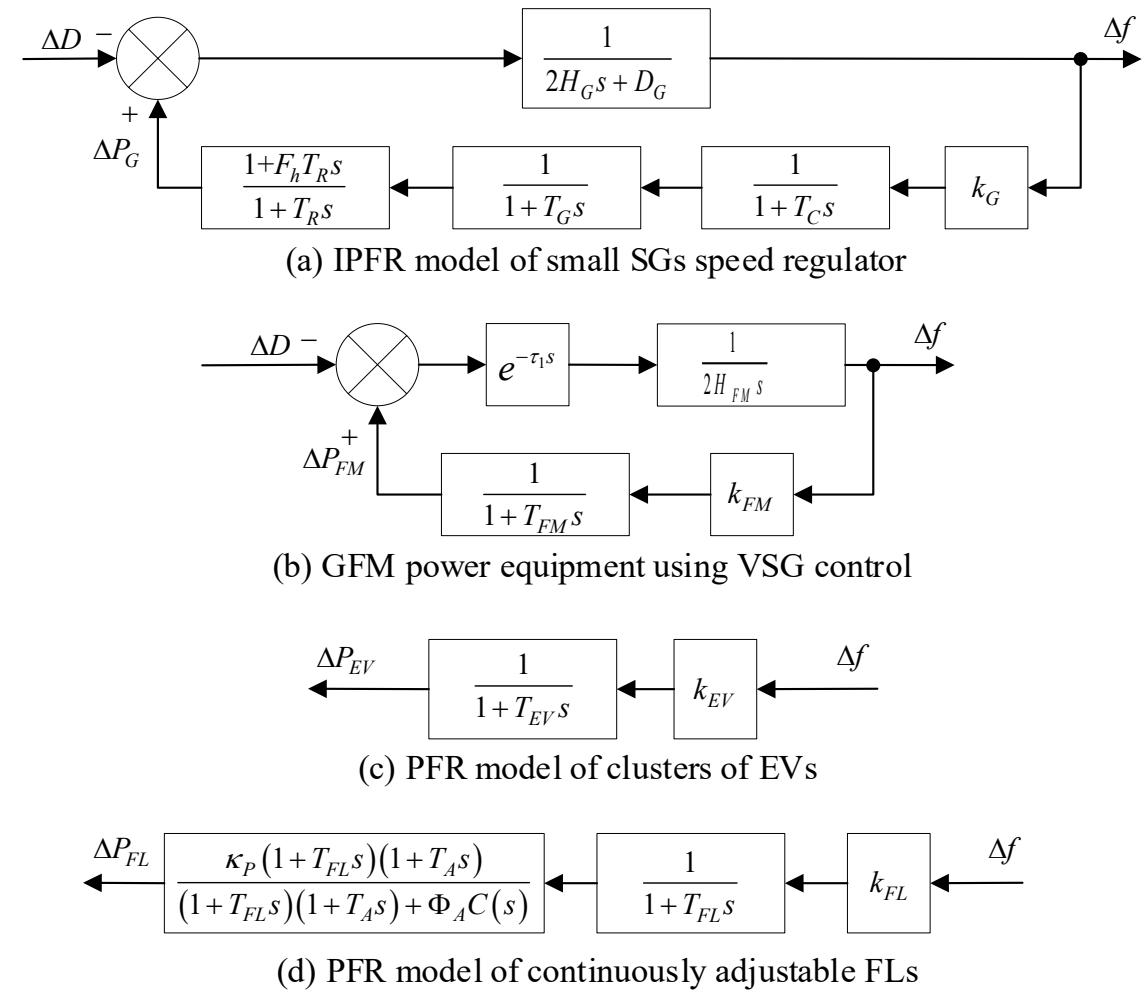

Fig. 3. IPFR model of different types of DERs.

### *B. Frequency Response Aggregation of Massive DERs*

VPPs integrate heterogeneous DER portfolios, necessitating the coordinated aggregation of inertial responses as characterized in Fig.3. Given the impact of inertial response delays on the IPFR market, the inertial response—primarily from small SGs and GFM IBRs—is categorized as either non-delayed or delayed. The aggregated inertia of the VPP is then determined as follows:

$$\begin{cases} H_{VPPG} = \sum\limits_{v \in N_G} H_{G,v} K_{m,v} \\ H_{VPPR} = \sum\limits_{v \in N_{FM}} H_{FM,v} K_{m,v} \end{cases} \tag{3}$$

where $K_{m,v}=S_v/S_{sys}$ represents the ratio of the rated power of unit $v$ to the total system power.

For homogeneous DERs (e.g., small SG #1, small SG #2), the frequency response models share an identical structure, differing only in their parameter values. Consequently, a weighted average method is employed to aggregate these models into a single equivalent representation. This aggregation process is illustrated in Fig. 4, using small SGs as an example to demonstrate the consolidation of a multi-machine model into a simplified single-machine model.

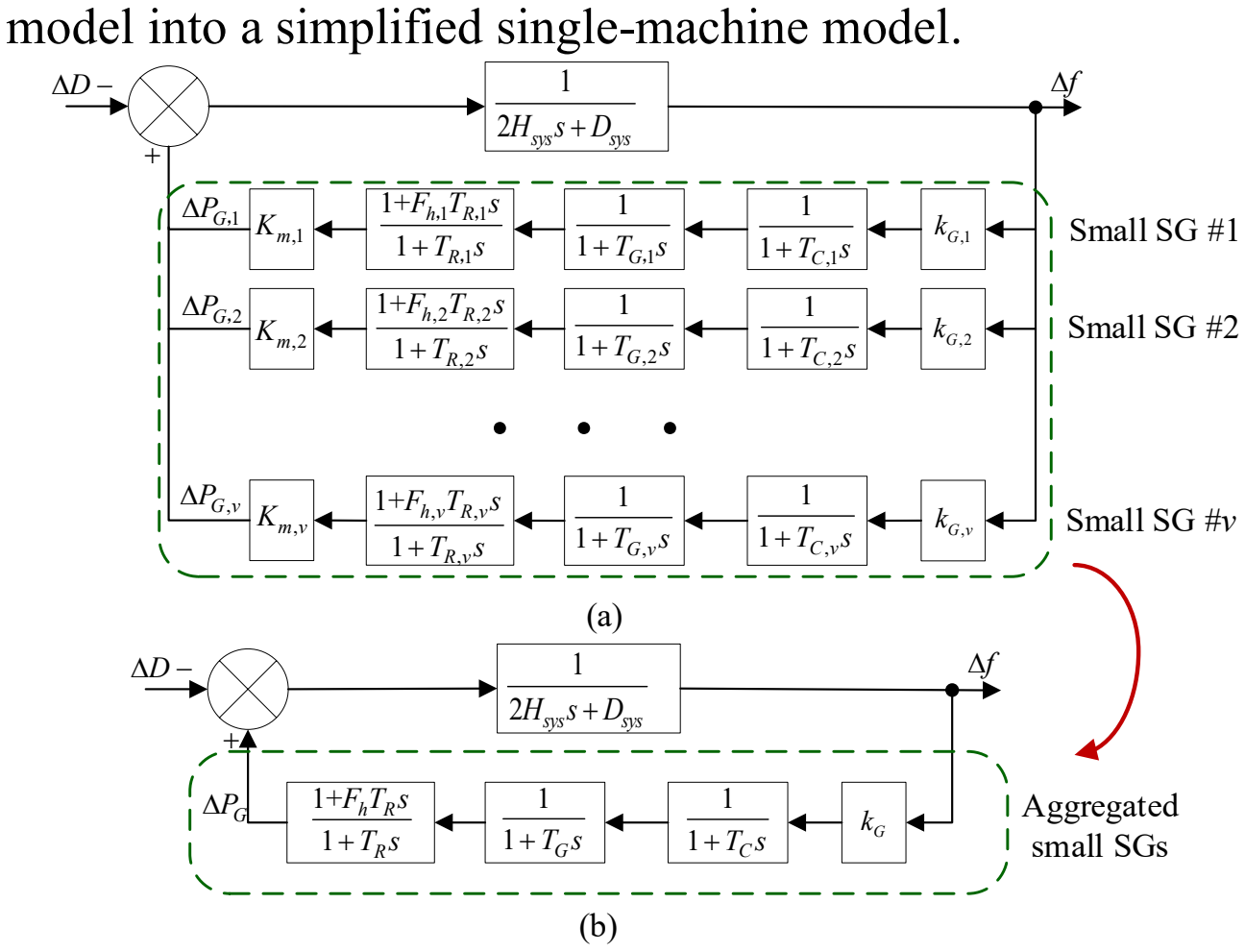

Fig. 4. Aggregated frequency response model of homogeneous machines.

In Fig. 4, the equivalent droop factor can be calculated as:

$$k_G = \sum_{v \in N_G} K_{m,v} k_{G,v} = K_G \sum_{v \in N_G} k_{G,v} \tag{4}$$

A normalized gain $\lambda_v = K_{m,v}k_{G,v}/k_G$ is defined for each feedback loop, which enables the homogeneous multi-machine transfer function to be aggregated into a single-machine model with the following parameters:

$$\begin{cases} F_h = \sum_{v\in N_G} \lambda_v F_{h,v}, T_R = \sum_{v\in N_G} \lambda_v T_{R,v} \\ T_G = \sum_{v\in N_G} \lambda_v T_{G,v}, T_C = \sum_{v\in N_G} \lambda_v T_{C,v} \end{cases} \quad (5)$$

The simplification process and its theoretical proof are detailed in [25].

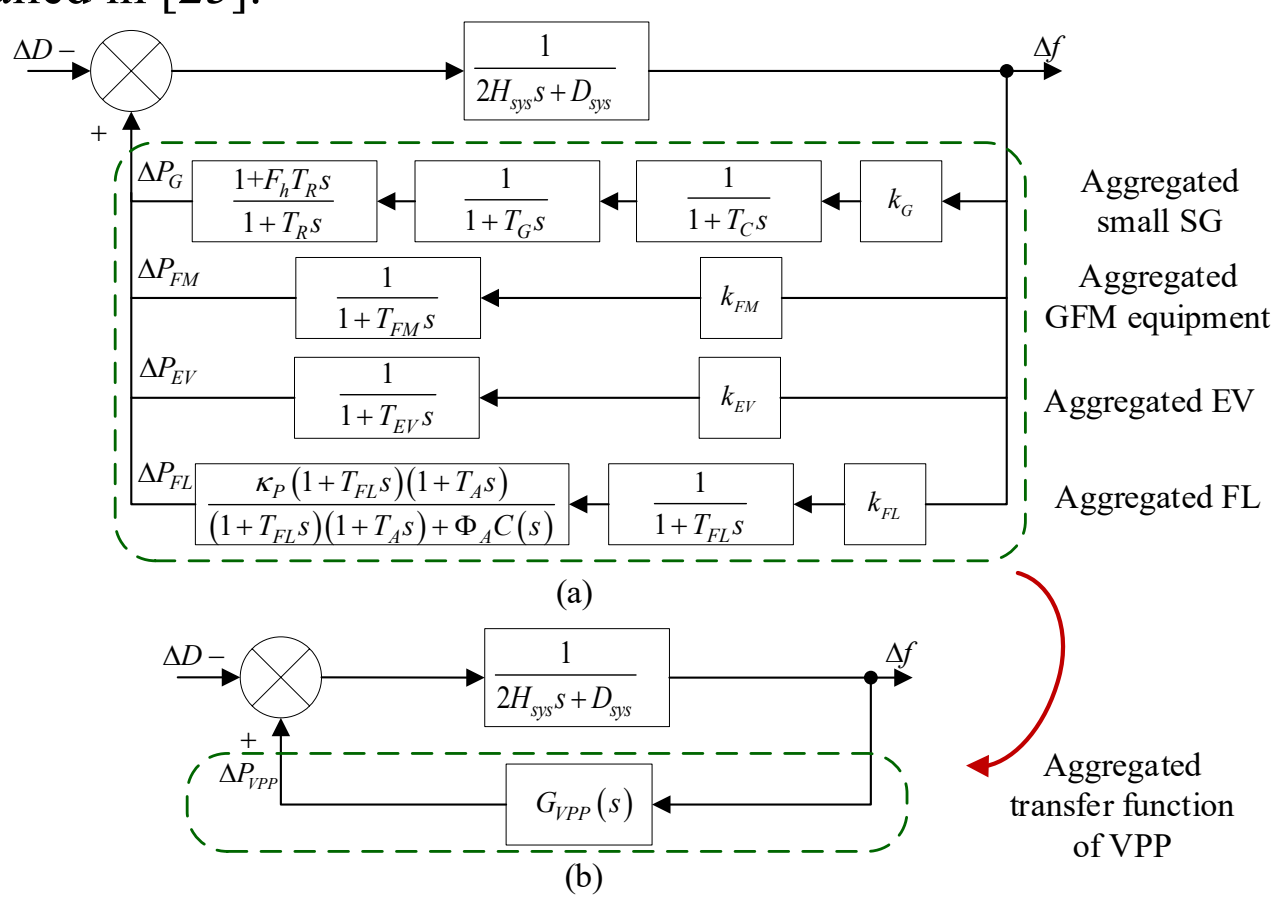

Fig. 5. Frequency response model of VPP considering multiple types of DERs.

As shown in Fig. 5(a), the frequency response model of the VPP with heterogeneous DERs can be derived using the weighted average method. To enable the VPP to provide a standardized PFR product in the IPFR market, an aggregated transfer function with a droop factor is proposed to describe its PFR feedback loop, which can be expressed as follows:

$$G_{VPP}(s) = \frac{k_\upsilon s^\upsilon + k_{\upsilon-1}s^{\upsilon-1} + \cdots + k^{VPP}}{h_\omega s^\omega + h_{\omega-1}s^{\omega-1} + \cdots + h_1 s + 1} \quad (6)$$

As indicated in (6), the aggregated multi-machine model in Fig. 5(a) can be further simplified into a single-input single-output (SISO) equivalent system whose transfer function contains only a small number of parameters depending on the chosen polynomial order. Thus, the aggregation method is scalable, with its compact SISO equivalent model avoiding parameter explosion despite increasing DER diversity.

### *C. Parameter Optimization of the VPP Aggregation Model*

To accurately represent the VPP's PFR dynamics, the aggregated transfer function $G_{VPP}(s)$ is designed to capture key performance criteria, specifically the frequency nadir ($f_{\text{nadir}}$) and the QSS frequency ($f_{\text{qss}}$), which represent the most severe dip and the settling point after a disturbance. The specific criteria for the multi-machine model and aggregation model are given in (7).

$$\begin{cases} f_{nadir} = f_0 + \Delta f(t_{nadir}), f_{qss} = f_0 + \Delta f(\infty) \\ \tilde{f}_{nadir} = f_0 + \Delta\tilde{f}(t^*_{nadir}), \tilde{f}_{qss} = f_0 + \Delta\tilde{f}(\infty) \end{cases} \quad (7)$$

Thus, the parameters of the aggregation model can be determined by solving the optimization problem in (8)-(9).

$$\min_{\Delta D} \mathbb{E}\left[\left(\Delta\tilde{f}(t^*_{nadir}) - \Delta f(t_{nadir})\right)^2 + \left(\Delta\tilde{f}(\infty) - \Delta f(\infty)\right)^2\right] \quad (8)$$

$$\text{s.t.} \quad \begin{cases} \Delta f'(t_{nadir}) = 0, \Delta\tilde{f}'(t^*_{nadir}) = 0 \\ \Delta f'(\infty) = 0, \Delta\tilde{f}'(\infty) = 0 \end{cases} \quad (9)$$

The objective function in (8) minimizes the errors in both $f_{\text{nadir}}$ and $f_{\text{qss}}$. This optimization is subject to the constraint defined in (9), which ensures consistency between the aggregated and multi-machine models. Given that the disturbance is random, the expectation is used with the predicted distribution of $\Delta D$ offered by the SO.

It should be noted that the constraints in (9) are non-convex due to the complex nonlinear relationship between the aggregated parameters and the system frequency dynamics. Although equality-constrained variants of SGD are available, they offer no convergence advantage for non-convex problems and introduce significant computational complexity. Therefore, the constrained optimization problem, i.e., (8)-(9), is relaxed using the penalty function method, and the resulting formulation can be efficiently solved using the SGD algorithm [36] with computational efficiency, simplicity, and convergence guarantees.

Since $t^*_{\text{nadir}}$ can be calculated from the equation $\Delta f'(t^*_{\text{nadir}}) = 0$ directly, $t^*_{\text{nadir}}$, $\Delta f(t^*_{\text{nadir}})$, and $\Delta f(\infty)$ are treated as known quantities and only the constraints $\Delta\tilde{f}'(t^*_{\text{nadir}}) = 0$ and $\Delta\tilde{f}'(\infty) = 0$ are penalized. Consequently, the optimization problem in (8)-(9) is reformulated into (10).

$$\begin{aligned} \min_{\Delta D} \mathbb{E}\Big[&\left(\Delta\tilde{f}(t_{nadir}) - \Delta f(t_{nadir})\right)^2 + \left(\Delta\tilde{f}(\infty) - \Delta f(\infty)\right)^2\Big] \\ &+ \lambda_{nadir}\left\|\Delta\tilde{f}'(t_{nadir})\right\|^2 + \lambda_{qss}\left\|\Delta\tilde{f}'(\infty)\right\|^2 \end{aligned} \quad (10)$$

In (10), a larger $\lambda_{\text{nadir}}$ improves the accuracy of the timing of the $f_{\text{nadir}}$, and a larger $\lambda_{\text{qss}}$ guarantees the steady-state condition is met. Thereby, the model upholds physical consistency, which is essential to prevent the aggregated droop factor from deviating from its practical basis. A parameter sweep across various disturbance scenarios revealed that values of $\lambda_{\text{nadir}}$ and $\lambda_{\text{qss}}$ in $[10^3, 10^4]$ consistently minimized the mean absolute percentage error (MAPE) for both $f_{\text{nadir}}$ and $f_{\text{qss}}$.

## IV. Dynamic Frequency Response of System with VPP

In this section, the analytical formula for the multi-stage system frequency dynamics incorporating response delays is derived. Based on this formula, the frequency-security constraints are then established for the IPFR market.

### *A. IPFR Dynamics Considering Frequency Response Delays*

SGs, GFM IBRs, and VPPs are considered participants in the IPFR market. For SGs, the IPFR dynamics are described by (1) and (2), while the virtual inertia and droop factor provision from GFM IBRs have been established in Section III-A. For brevity, the time constant of the GFM inverter is neglected when deriving the frequency response of GFM IBRs, as its value is nearly milliseconds, much smaller than the time scales of interest for system frequency stability [33]. The corresponding expressions are given below:

$$\Delta P_k^{ESS} = -2H_k^{ESS} \cdot \frac{d\Delta f(t)}{dt} - k_k^{ESS} \cdot \Delta f(t), k \in \Omega_{ESS}, t \geq \tau_1 \quad (11)$$

$$\Delta P_j^{REG} = -2H_j^{REG} \cdot \frac{d\Delta f(t)}{dt} - k_j^{REG} \cdot \Delta f(t), j \in \Omega_{REG}, t \geq \tau_1 \quad (12)$$

In this study, a first-order transfer function, i.e., $G_{VPP}(s) = k^{VPP}/(1+T^{VPP}s)$, is used to represent the PFR feedback loop of the aggregated VPP. Thus, the frequency dynamics of IPFR provided by VPP can be expressed as (13) and (14).

$$\begin{cases} 2H_l^{VPPG} \cdot \dfrac{d\Delta f(t)}{dt} = \Delta P_l^{VPPR} - \Delta D, l \in \Omega_{VPP}, t < \tau_1 \\ 2H_l^{VPP} \cdot \dfrac{d\Delta f(t)}{dt} = \Delta P_l^{VPP} - \Delta D, l \in \Omega_{VPP}, t \geq \tau_1 \end{cases} \quad (13)$$

$$T_l^{VPP} \cdot \frac{d\Delta P_{t,l}^{VPP}}{dt} + \Delta P_{t,l}^{VPP} = -k_l^{VPP} \cdot \Delta f(t), l \in \Omega_{VPP} \quad (14)$$

Given the diverse IPFR delays in different resources, the system's response to a disturbance (at $t$=$0^+$) can be divided into three stages:

Stage 1 ($0 \leq t \leq \tau_1$): Inertial response is provided by online SGs participating directly in the market and small SGs within VPPs.

Stage 2 ($\tau_1 < t \leq \tau_2$): GFM IBRs and converter-interfaced DERs within VPPs begin to deliver virtual inertia and droop factors.

Stage 3 ($t > \tau_2$): The PFR from online SGs and small SGs within VPPs is activated.

Thus, the system frequency dynamics can be expressed as:

$$\begin{cases} 2H^{GV} \cdot \dfrac{d\Delta f(t)}{dt} = -\Delta D, 0 \leq t \leq \tau_1 \\ 2H^{TO} \cdot \dfrac{d\Delta f(t)}{dt} = \Delta P^{FM} + \Delta P^{VPPR} - \Delta D, \tau_1 < t \leq \tau_2 \\ 2H^{TO} \cdot \dfrac{d\Delta f(t)}{dt} = \Delta P^{G} + \Delta P^{FM} + \Delta P^{VPP} - \Delta D, t > \tau_2 \end{cases} \quad (15)$$

where:

$$\begin{cases} H^{GV} = \sum_{i \in \Omega_G} x_{i,t} H_i^G + \sum_{l \in \Omega_{VPP}} x_{l,t} H_l^{VPPG} \\ H^{TO} = H^{GV} + \sum_{j \in \Omega_{REG}} H_j^{REG} + \sum_{k \in \Omega_{ESS}} H_k^{ESS} + \sum_{l \in \Omega_{VPP}} H_l^{VPPR} \\ \Delta P^{FM} = \sum_{k \in \Omega_{ESS}} \Delta P_k^{ESS} + \sum_{j \in \Omega_{REG}} \Delta P_j^{REG} \end{cases} \quad (16)$$

### B. Frequency-security Constraints

Based on the process of multi-stage frequency response, frequency-security constraints are formulated according to the metrics mentioned in Section II-A. Although the IPFR from GFM IBRs is subject to a delay $\tau_1$ and thus does not contribute to the instantaneous inertial response, this delay $\tau_1$ (0.01-0.05 s , see [38]) is negligible compared to the PFR delay $\tau_2$ of SGs (1-3 s, see [39]) when deriving frequency behavior for $t > 0^+$. By combining (2) and (11)-(16) under the initial conditions $\Delta f|_{t=0+} = 0$, the analytical formula for system frequency can be derived as (17).

$$\begin{cases} \Delta f_1(t) = \dfrac{\Delta D}{2H^{GV}} t, t=0^+ \\ \Delta f_2(t) = \dfrac{\Delta D}{k^{FM} + k^{VPPR}} (e^{-\frac{k^{FM} + k^{VPPR}}{2H^{TO}} t} - 1), 0^+ < t \leq \tau_2 \\ \Delta f_3(t) = \Delta D' e^{-\alpha t} [C_1 \sin(\omega t) + C_2 \cos(\omega t)] \\ \qquad - \dfrac{\Delta D'}{k^G + k^{FM} + k^{VPP}}, t > \tau_2 \end{cases} \quad (17)$$

where $\alpha$, $\omega$, $\Delta D'$, $C_1$, and $C_2$ are parameters of the dynamic analytical formula. The detailed derivation of (17) is provided in Appendix A.

Using the formula (17), the following frequency-security constraints are established:

1) RoCoF constraint: The maximum RoCoF occurs at the beginning of the disturbance, when the power imbalance is maximal. Thus, by substituting $t$=$0^+$ into (17), it yields:

$$RoCoF_{\max} \geq \frac{\Delta D}{2H^{GV}} \quad (18)$$

2) Frequency nadir constraint: Since $\Delta f_2(t)$ decreases monotonically, the nadir frequency occurs when $\Delta f_3'(t_{\text{nadir}})$=0 for $t > \tau_2$. Substituting initial conditions into (17), it leads to a nonlinear constraint, which is reformulated by piecewise linearization (PWL) for tractability [17] as shown in (19).

$$\begin{aligned} \kappa_{hp}^1 H^{TO} + \kappa_{hp}^2 k^G + \kappa_{hp}^3 k^{FM} + \kappa_{hp}^4 k^{VPP} + \kappa_{hp}^5 \\ \geq -\Delta f_{nadir-\max}, hp \in N_{hp} \end{aligned} \quad (19)$$

3) QSS frequency constraint: By substituting $t \to \infty$ into (17), it yields:

$$\Delta f_{ss}^{ref} \geq \frac{\Delta D}{k^G + k^{FM} + k^{VPP}} \quad (20)$$

## V. Market-clearing Model

In this section, a joint energy-reserve-IPFR market-clearing model is designed considering REGs' generation uncertainty and distinct IPFR characteristics of SGs, GFM IBRs (i.e., GFM inverter-based REGs and GFM inverter-based ES units), and VPPs.

### A. Framework for Market-clearing Model

A unit commitment (UC) and economic dispatch (ED) model, incorporating REGs' generation uncertainty as well as inertia and droop factor provisions, are developed to clear the joint energy–reserve–IPFR market. The overall structure of the market-clearing process is outlined in Fig. 6. A security-constrained unit commitment (SCUC) model and its continuous relaxation, denoted as the continuous SCUC model, are solved in parallel. In this process, the SCUC model determines the UC of SGs, while the continuous SCUC model yields marginal prices for inertia and droop factors provided by SGs [40]. Subsequently, with the SG commitment fixed, a security-constrained economic dispatch (SCED) model is used to optimally allocate and price energy, reserve, inertia, and droop factors for GFM IBRs and VPPs. Additionally, the REGs' generation uncertainty is addressed by chance constraints (CCs).

The proposed clearing model achieves strong duality and interpretable marginal pricing through the two convexified measures:

1) Reformulation of non-convex constraints: CCs are transformed into tractable deterministic equivalents, and the nonlinear frequency nadir constraints are approximated using PWL.

2) Convex relaxation of binary variables: The start-up and shut-down statuses of small SGs within VPPs are represented as continuous variables, reflecting their operational flexibility and short start-up/shut-down cycles (less than one hour, see [37]). Moreover, the binary commitment variables are also relaxed in the continuous SCUC model.

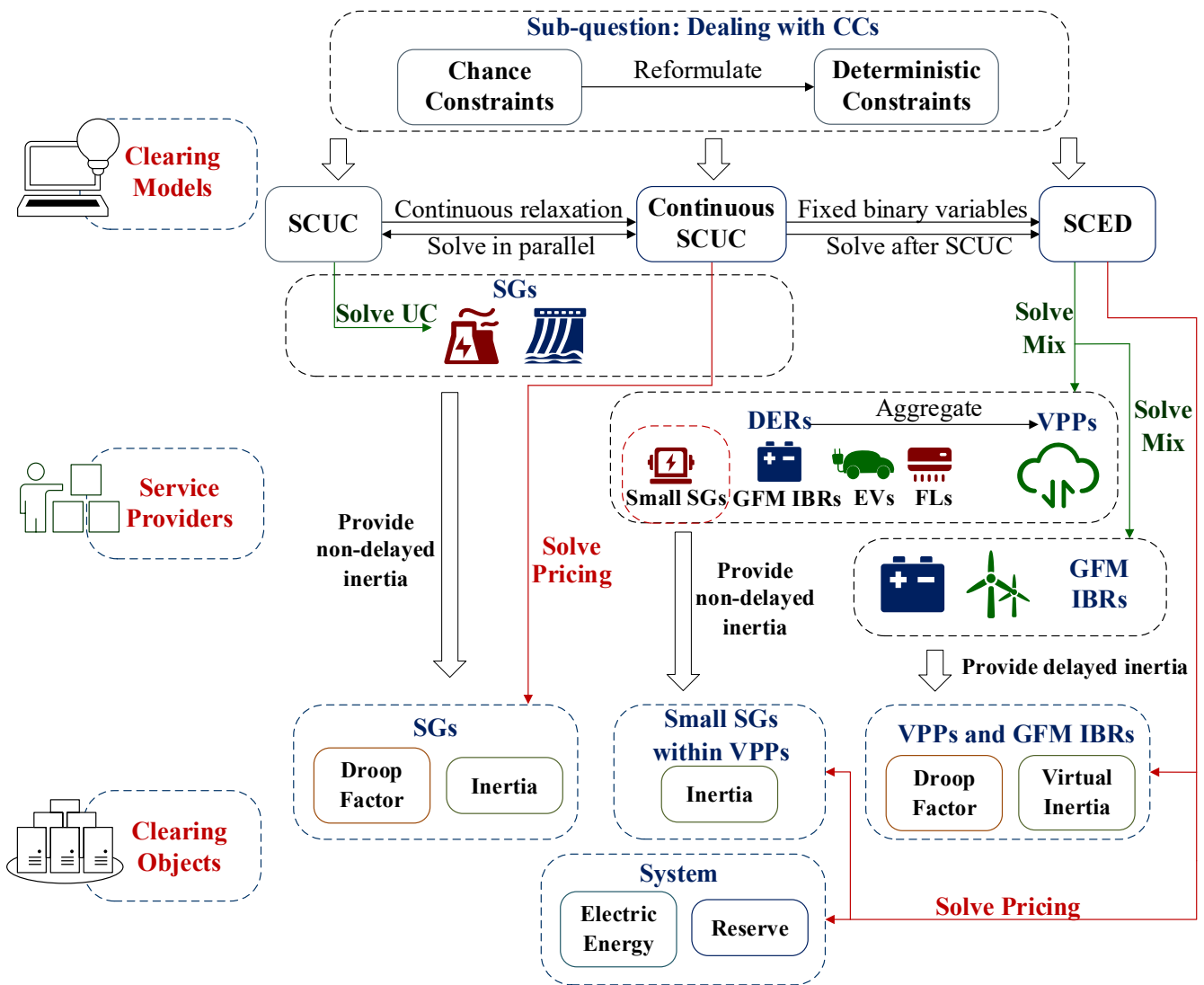

Fig. 6. Solving structure of the joint energy-reserve-IPFR market-clearing model.

These constructions ensure that the resulting optimization problem is convex, allowing the clearing prices for all services to be directly and uniquely derived from the dual variables of their respective constraints. This guarantees a transparent and resource-specific marginal pricing scheme.

## B. Stochastic Modeling and Balancing Regulation for REG

**Bold** symbols denote random variables hereafter, except noted. The stochastic output power of GFM inverter-based REG $j$ at period $t$ is defined as $\boldsymbol{P}_{j,t}^{REG,EN}$, which is composed of a deterministic forecast power $P_{j,t}^{REG,EN}$ and a random forecast error $\boldsymbol{\varepsilon}_{j,t}$. It is assumed that the forecast error is distributed as $\boldsymbol{\varepsilon}_{j,t} \sim N(\mu_{j,t}, \sigma_{j,t}^2)$, where $\mu_{j,t}$ and $\sigma_{j,t}^2$ denote the mean and variation, respectively [41]. Hence, it follows that:

$$\boldsymbol{P}_{j,t}^{REG,EN} = P_{j,t}^{REG,EN} - \boldsymbol{\varepsilon}_{j,t} \tag{21}$$

The system-wide forecast error of REGs is defined as $\boldsymbol{\varepsilon}_{sys,t} = e^{\top}\boldsymbol{\varepsilon}_t$, where $\boldsymbol{\varepsilon}_t$ is a column vector summing all the forecast errors in period $t$, and $e$ is a column vector of ones with the appropriate dimension. Thus, $\boldsymbol{\varepsilon}_{sys,t}$ follows a normal distribution, i.e., $\boldsymbol{\varepsilon}_{sys,t} \sim N(\mu_{sys,t}, \sigma_{sys,t}^2)$. The forecast error $\boldsymbol{\varepsilon}_{sys,t}$ is compensated in the reserve market by SGs, ES units, and VPPs, with balancing participation factors $u_{i,t}$, $u_{k,t}^{c/d}$, $u_{l,t} \in [0, 1]$, which determine their real-time power outputs as described in (22)-(25).

$$\boldsymbol{P}_{i,t}^{G} = P_{i,t}^{G} + u_{i,t}\boldsymbol{\varepsilon}_{sys,t} \tag{22}$$

$$\boldsymbol{P}_{k,t}^{ESS,d} = P_{k,t}^{ESS,d} + u_{k,t}^{d}\boldsymbol{\varepsilon}_{sys,t} \tag{23}$$

$$\boldsymbol{P}_{k,t}^{ESS,c} = P_{k,t}^{ESS,c} - u_{k,t}^{c}\boldsymbol{\varepsilon}_{sys,t} \tag{24}$$

$$\boldsymbol{P}_{l,t}^{VPP,EN} = P_{l,t}^{VPP,EN} + u_{l,t}\boldsymbol{\varepsilon}_{sys,t} \tag{25}$$

It is worth noting that the assumption of a normal distribution in this study is primarily for illustrative purposes. The proposed framework is not restricted to any specific probability distribution and can accommodate other distributions by adjusting the chance-constraint reformulations accordingly.

## C. Objective Function of the Clearing Model

The objective function of the joint energy-reserve-IPFR market is to minimize the social generation cost, which consists of electric energy cost, inertia cost, droop factor cost, start-up cost, and shut-down cost. These costs are described by (26)-(31).

$$C^{TO} = \min \sum_{t=1}^{N_T} \left( c_t^{EN} + c_t^{IN} + c_t^{DF} + c_t^{SU} + c_t^{SD} \right) \Delta t \tag{26}$$

$$\begin{aligned} c_t^{EN} = {} & \mathbb{E}_{\boldsymbol{\varepsilon}_t}\left[ \sum_{i\in\Omega_G} \alpha_{i,t}^{G} \boldsymbol{P}_{i,t}^{G} \right] + \sum_{j\in\Omega_{REG}} \alpha_{j,t}^{REG} P_{j,t}^{REG,EN} \\ & + \mathbb{E}_{\boldsymbol{\varepsilon}_t}\left[ \sum_{k\in\Omega_{ESS}} \alpha_{k,t}^{ESS} \left( \boldsymbol{P}_{k,t}^{ESS,c} - \boldsymbol{P}_{k,t}^{ESS,d} \right) + \sum_{l\in\Omega_{VPP}} \alpha_{l,t}^{VPP} \boldsymbol{P}_{l,t}^{VPP,EN} \right] \end{aligned} \tag{27}$$

$$c_t^{IN} = \sum_{j\in\Omega_{REG}} \beta_{j,t}^{REG} H_{j,t}^{REG} + \sum_{k\in\Omega_{ESS}} \beta_{k,t}^{ESS} H_{k,t}^{ESS} + \sum_{l\in\Omega_{VPP}} \beta_{l,t}^{VPPR} H_{l,t}^{VPPR} \tag{28}$$

$$c_t^{DF} = \sum_{j\in\Omega_{REG}} \chi_{j,t}^{REG} k_{j,t}^{REG} + \sum_{k\in\Omega_{ESS}} \chi_{k,t}^{ESS} k_{k,t}^{ESS} + \sum_{l\in\Omega_{VPP}} \chi_{l,t}^{VPP} k_{l,t}^{VPP} \tag{29}$$

$$c_t^{SU} = \sum_{i\in\Omega_G} \left( y_{i,t}^{SU} C_i^{SU-N} + h_{i,t}^{SU} C_i^{SU-M} \right) \tag{30}$$

$$c_t^{SD} = \sum_{i\in\Omega_G} y_{i,t}^{SD} C_i^{SD} \tag{31}$$

As warm and hot starts are widely adopted among SGs, a linear function is employed to model the start-up cost in (30), which also guarantees sufficient accuracy [42]. However, the inclusion of start-up and shut-down costs in the clearing model may lead to huge fluctuations in marginal prices due to their high and concentrated nature. To address this, the offline hours $h_{i,t}^{SU}$ are modeled as the cumulative sum of shut-down states, i.e., $\Sigma_{t'\in\mathcal{T}}(1 - x_{i,t'})$, and thus the corresponding start-up costs are allocated across the relevant offline periods within the set $\mathcal{T}$.

## D. Constraints of the Clearing Model

The constraints of the clearing model mainly consist of operational constraints for the power system and all service providers (e.g., SGs, GFM IBRs, and VPPs).

### 1) Power System Constraints

$$\begin{aligned} & \sum_{i\in\Omega_G} P_{i,t}^{G} + \sum_{j\in\Omega_{REG}} P_{j,t}^{REG,EN} + \sum_{k\in\Omega_{ESS}} P_{k,t}^{ESS,EN} + \sum_{l\in\Omega_{VPP}} P_{l,t}^{VPP,EN} \\ & - \sum_{n\in\Omega_n} D_{n,t} - \sum_{m\in\Omega_n^N} B_{n,m}(\theta_{n,t} - \theta_{m,t}) = 0, \forall n, \forall t : \lambda_{n,t}^{EN} \end{aligned} \tag{32}$$

$$\sum_{i\in\Omega_G} u_{i,t} + \sum_{k\in\Omega_{ESS}} (u_{k,t}^{d} - u_{k,t}^{c}) + \sum_{l\in\Omega_{VPP}} u_{l,t} = 1, \forall t : \lambda_t^{R} \tag{33}$$

$$H_t^{GV} \geq \frac{\Delta D}{2\left| RoCoF_{\max} \right|}, \forall t : \lambda_t^{IN} \tag{34}$$

$$k_t^{G} + k_t^{FM} + k_t^{VPP} \geq \frac{\Delta D}{\left| \Delta f_{ss}^{ref} \right|}, \forall t : \lambda_t^{DR} \tag{35}$$

$$\kappa_{hp}^{1}H^{TO}+\kappa_{hp}^{2}k^{G}+\kappa_{hp}^{3}k^{FM}+\kappa_{hp}^{4}k^{VPP}+\kappa_{hp}^{5} \geq -\Delta f_{nadir-\max}, hp \in N_{hp}, \forall t : \lambda_{t}^{IPFR} \quad (36)$$

Cons. (32) ensures system-wide power balance while Cons. (33) ensures the reserve sufficiency. Additionally, Cons. (34)-(36) outline the frequency-security constraints derived in Section IV-B. The dual multipliers for Cons. (32)-(36) are denoted as $\lambda_{n,t}^{EN}$, $\lambda_{t}^{R}$, $\lambda_{t}^{IN}$, $\lambda_{t}^{DR}$, and $\lambda_{hp,t}^{IPFR}$, respectively.

*2) SG Operating Constraints*

$$\mathbb{P}_{\boldsymbol{\varepsilon}_t}\left[x_{i,t}P_{i,t,\min}^{G,EN} \leq \boldsymbol{P}_{i,t}^{G} \leq x_{i,t}P_{i,t,\max}^{G,EN}\right] \geq 1-p_{i}^{G}, \forall i \in \Omega_{G}, \forall t \quad (37)$$

$$\mathbb{P}_{\boldsymbol{\varepsilon}_t}\left[-P_{i}^{Ramp} \leq \boldsymbol{P}_{i,t}^{G}-\boldsymbol{P}_{i,t-1}^{G} \leq P_{i}^{Ramp}\right] \geq 1-p_{i}^{G}, \forall i \in \Omega_{G}, \forall t \quad (38)$$

$$x_{i,t}-x_{i,t-1}=y_{i,t}^{SU}-y_{i,t}^{SD}, \forall i \in \Omega_{G}, \forall t \quad (39)$$

Cons. (37) limits the power output of each SG within its feasible operating range, while Cons. (38) enforces ramping limits. Cons. (39) models the logical relationship between the start-up and shut-down statuses.

*3) GFM Inverter-based REG Operation Constraints*

$$P_{j,\min}^{REG,EN} \leq P_{j,t}^{REG,EN} \leq P_{j,\max}^{REG,EN}, \forall j \in \Omega_{REG}, \forall t \quad (40)$$

$$P_{j,\min}^{REG,IN} \leq \underbrace{2\left|RoCoF_{\max}\right| \cdot H_{j,t}^{REG}}_{P_{j,t}^{REG,IN}} \leq P_{j,\max}^{REG,IN}, \forall j \in \Omega_{REG}, \forall t \quad (41)$$

$$P_{j,\min}^{REG,DF} \leq \underbrace{(\left|\Delta f_{\max-nadir}\right| \cdot k_{j,t}^{REG})}_{P_{j,t}^{REG,DF}} \leq P_{j,\max}^{REG,DF}, \forall j \in \Omega_{REG}, \forall t \quad (42)$$

$$P_{j,t}^{REG,EN}+P_{j,t}^{REG,IN}+P_{j,t}^{REG,DF} \leq P_{j,t}^{REG}, \forall j \in \Omega_{REG}, \forall t \quad (43)$$

Cons. (40)-(43) specify the operating limits for GFM inverter-based REGs, where $P_{j,t}^{REG,EN}$, $P_{j,t}^{REG,IN}$, and $P_{j,t}^{REG,DF}$ are treated as deterministic.

*4) GFM Inverter-based ES Operation Constraints*

$$\mathbb{P}_{\boldsymbol{\varepsilon}_t}\left[\boldsymbol{P}_{k,t}^{ESS,c} \leq z_{k}P_{k,\max}^{ESS,c}\right] \geq 1-p_{k}^{c}, \forall k \in \Omega_{ESS}, \forall t \quad (44)$$

$$\mathbb{P}_{\boldsymbol{\varepsilon}_t}\left[\boldsymbol{P}_{k,t}^{ESS,d} \leq \left(1-z_{k}\right)P_{k,\max}^{ESS,d}\right] \geq 1-p_{k}^{d}, \forall k \in \Omega_{ESS}, \forall t \quad (45)$$

$$S_{k,\min}^{ESS} \leq S_{k,t}^{ESS} \leq S_{k,\max}^{ESS}, \forall k \in \Omega_{ESS}, \forall t \quad (46)$$

$$S_{k,t}^{ESS}=S_{k,t-1}^{ESS}+\mathbb{E}_{\boldsymbol{\varepsilon}_t}\left[\mu_{k}^{c}\boldsymbol{P}_{k,t-1}^{ESS,c}-\boldsymbol{P}_{k,t-1}^{ESS,d}/\mu_{k}^{d}\right], \forall k \in \Omega_{ESS}, \forall t \quad (47)$$

$$P_{k,\min}^{ESS,IN} \leq \underbrace{2\left|RoCoF_{\max}\right| \cdot H_{k,t}^{ESS}}_{P_{k,t}^{ESS,IN}} \leq P_{k,\max}^{ESS,IN}, \forall k \in \Omega_{ESS}, \forall t \quad (48)$$

$$P_{k,\min}^{ESS,DF} \leq \underbrace{(\left|\Delta f_{\max-nadir}\right| \cdot k_{k,t}^{ESS})}_{P_{k,t}^{ESS,DF}} \leq P_{k,\max}^{ESS,DF}, \forall k \in \Omega_{ESS}, \forall t \quad (49)$$

$$0 \leq (P_{k,t}^{ESS,IN}+P_{k,t}^{ESS,DR})\Delta t^{IPFR} \leq S_{k,t}^{ESS}-S_{k,\min}^{ESS}, \forall k \in \Omega_{ESS}, \forall t \quad (50)$$

$$\mathbb{P}_{\boldsymbol{\varepsilon}_t}\left[\boldsymbol{P}_{k,t}^{ESS,d}+P_{k,t}^{ESS,IN}+P_{k,t}^{ESS,DF} \leq P_{k,\max}^{ESS,d}\right] \geq 1-p_{k}^{d}, \forall k \in \Omega_{ESS}, \forall t \quad (51)$$

Cons. (44) and (45) impose CCs on the charging/ discharging power of ES units. Cons. (46) and (47) bound the state of charge. Cons. (48)-(51) ensure that sufficient energy capacity is always available for frequency response services.

*5) VPP Operation Constraints*

$$\mathbb{P}_{\boldsymbol{\varepsilon}_t}\left[x_{l,t}K_{G}P_{l,t,\min}^{VPP} \leq \boldsymbol{P}_{l,t}^{VPPG} \leq x_{l,t}K_{G}P_{l,t,\max}^{VPP}\right] \geq 1-p_{l}^{VPP}, \forall l \in \Omega_{VPP}, \forall t \quad (52)$$

$$\mathbb{P}_{\boldsymbol{\varepsilon}_t}\left[-P_{l}^{Ramp} \leq \boldsymbol{P}_{l,t}^{VPPG}-\boldsymbol{P}_{l,t-1}^{VPPG} \leq P_{l}^{Ramp}\right] \geq 1-p_{l}^{VPP}, \forall l \in \Omega_{VPP}, \forall t \quad (53)$$

$$\mathbb{P}_{\boldsymbol{\varepsilon}_t}\left[0 \leq \boldsymbol{P}_{l,t}^{VPPR,EN} \leq \left(1-K_{G}\right)P_{l,t,\max}^{VPP}\right] \geq 1-p_{l}^{VPP}, \forall l \in \Omega_{VPP}, \forall t \quad (54)$$

$$P_{l,\min}^{VPPR,IN} \leq \underbrace{2\left|RoCoF_{\max}\right| \cdot H_{l,t}^{VPPR}}_{P_{l,t}^{VPPR,IN}} \leq P_{l,\max}^{VPPR,IN}, \forall l \in \Omega_{VPP}, \forall t \quad (55)$$

$$P_{l,\min}^{VPP,DF} \leq \underbrace{(\left|\Delta f_{\max-nadir}\right| \cdot k_{l,t}^{VPP})}_{P_{l,t}^{VPP,DF}} \leq P_{l,\max}^{VPP,DF}, \forall l \in \Omega_{VPP}, \forall t \quad (56)$$

$$P_{l,t}^{VPP,EN}=P_{l,t}^{VPPG}+P_{l,t}^{VPPR,EN}, \forall l \in \Omega_{VPP}, \forall t \quad (57)$$

Cons. (52) and (53) maintain the power output and ramping limits of small SGs within the VPPs. Cons. (54) restricts the power output of converter-interfaced DERs within the VPPs. Cons. (55) and (56) limit the capacity provision of virtual inertia and droop factor, respectively. Cons. (57) ensures the power balance of each VPP. Since the VPP is responsible for scheduling the affiliated DERs, the operational constraints of individual DERs within the VPP are not explicitly captured in the system-wide clearing model.

## E. Deterministic Equivalent of the Model

The market-clearing model contains expectation and probability operators, which need to be reformulated into deterministic formats as follows.

*1) Expected Cost*

The electric energy cost in (27) depends on the stochastic power injection of SGs, ES units, and VPPs. Its expectation at time $t$ is therefore evaluated as:

$$\begin{aligned} c_{t}^{EN} &= \sum_{i \in \Omega_{G}}\left(\alpha_{i,t}^{G}P_{i,t}^{G}+\mu_{sys,t}u_{i,t}\right) \\ &+ \sum_{k \in \Omega_{ESS}}\left(\alpha_{k,t}^{ESS}P_{k,t}^{ESS,c}+\mu_{sys,t}u_{k,t}^{c}-\alpha_{k,t}^{ESS}P_{k,t}^{ESS,d}-\mu_{sys,t}u_{k,t}^{d}\right) \\ &+ \sum_{j \in \Omega_{REG}}\alpha_{j,t}^{REG}P_{j,t}^{REG,EN}+\sum_{l \in \Omega_{VPP}}\left(\alpha_{l,t}^{VPP}P_{l,t}^{VPP,EN}+\mu_{t,sys}u_{l,t}\right) \end{aligned} \quad (58)$$

*2) Chance Constraints*

By introducing the inverse cumulative distribution function $\Psi(p)$ of the system-wide forecast error $\boldsymbol{\varepsilon}_{\mathrm{sys},t}$, CCs, i.e., (37), (38), (44), (45), (47), and (51)-(54), are reformulated into their deterministic equivalents, yielding Cons. (59)-(67). This transformation leverages the quantile function of the forecast error distribution to replace each CC with a deterministic linear inequality, ensuring the prescribed violation probability is always met.

$$x_{i,t}P_{i,t,\min}^{G,EN}-\Psi\left(p_{i}^{G}\right)u_{i,t} \leq P_{i,t}^{G,EN} \leq x_{i,t}P_{i,t,\max}^{G,EN}+\Psi\left(p_{i}^{G}\right)u_{i,t}, \forall i \in \Omega_{G}, \forall t \quad (59)$$

$$-P_{i}^{Ramp}-\Psi\left(p_{i}^{G}\right)u_{i,t} \leq P_{i,t}^{G}-P_{i,t-1}^{G} \leq P_{i}^{Ramp}+\Psi\left(p_{i}^{G}\right)u_{i,t}, \forall i \in \Omega_{G}, \forall t \quad (60)$$

$$P_{k,t}^{ESS,c} \leq z_{k}P_{k,\max}^{ESS,c}+\Psi\left(p_{k}^{c}\right)u_{k,t}, \forall k \in \Omega_{ESS}, \forall t \quad (61)$$

$$P_{k,t}^{ESS,d} \leq \left(1-z_{k}\right)P_{k,\max}^{ESS,d}+\Psi\left(p_{k}^{d}\right)u_{k,t}, \forall k \in \Omega_{ESS}, \forall t \quad (62)$$

$$S_{k,t}^{ESS}=S_{k,t-1}^{ESS}+\mu_{k}^{c}\left(P_{k,t-1}^{ESS,c}+\mu_{sys,t}u_{k,t}^{c}\right)-\left(P_{k,t-1}^{ESS,d}+\mu_{sys,t}u_{k,t}^{d}\right)/\mu_{k}^{d}, \forall k \in \Omega_{ESS}, \forall t \quad (63)$$

$$P_{k,t}^{ESS,d}+P_{k,t}^{ESS,IN}+P_{k,t}^{ESS,DF}\le P_{k,\max}^{ESS,d}+\Psi\left(p_{k}^{d}\right)u_{k,t},\forall k\in\Omega_{ESS},\forall t \quad (64)$$

$$x_{l,t}K_{G}P_{l,t,\min}^{VPP}-\Psi\left(p_{l}^{VPP}\right)K_{G}u_{l,t}\le P_{l,t}^{VPPG}\le x_{l,t}K_{G}P_{l,t,\max}^{VPP}+\Psi\left(p_{l}^{VPP}\right)K_{G}u_{l,t},\forall l\in\Omega_{VPP},\forall t \quad (65)$$

$$-P_{l}^{Ramp}-\Psi\left(p_{l}^{VPP}\right)K_{G}u_{l,t}\le P_{l,t}^{VPPG}-P_{l,t-1}^{VPPG}\le P_{l}^{Ramp}+\Psi\left(p_{l}^{VPP}\right)K_{G}u_{l,t},\forall l\in\Omega_{VPP},\forall t \quad (66)$$

$$\Psi\left(p_{l}^{VPP}\right)K_{G}u_{l,t}-\Psi\left(p_{l}^{VPP}\right)u_{l,t}\le P_{l,t}^{VPPR,EN}\le\left(1-K_{G}\right)\left[P_{l,t,\max}^{VPP}+\Psi\left(p_{l}^{VPP}\right)u_{l,t}\right],\forall l\in\Omega_{VPP},\forall t \quad (67)$$

### *F. Pricing Mechanism*

Marginal prices for energy, reserve, inertia, and droop factors are derived from the dual variables of the proposed market-clearing model, leveraging strong duality and marginal-cost pricing principles. Specifically, the energy and reserve prices at bus $n$ and time $t$ are given by the multipliers $\lambda_{n,t}^{EN}$ and $\lambda_{t}^{R}$ associated with Cons. (32) and (33), respectively.

Marginal prices for inertia and droop factors are further differentiated by resource type and market-clearing stage. For SGs committed to the continuous SCUC model, the corresponding prices are extracted from the dual variables of Cons. (34)-(36), yielding (68)-(69).

$$\rho_{i,t}^{G,IN}=x_{i,t}\lambda_{t}^{IN}+x_{i,t}\sum_{hp\in N_{hp}}\kappa_{hp}^{1}\lambda_{hp,t}^{IPFR} \quad (68)$$

$$\rho_{i,t}^{G,DR}=x_{i,t}\lambda_{t}^{DR}+x_{i,t}\sum_{hp\in N_{hp}}\kappa_{hp}^{2}\lambda_{hp,t}^{IPFR} \quad (69)$$

Likewise, inertia and droop factor prices for GFM IBRs and VPPs in the SCED model are obtained from the dual variables of Cons. (34)-(36), yielding (70)-(73).

$$\rho_{t}^{FM,IN}=\sum_{hp\in N_{hp}}\kappa_{hp}^{1}\lambda_{hp,t}^{IPFR} \quad (70)$$

$$\rho_{t}^{FM,DR}=\lambda_{t}^{DR}+\sum_{hp\in N_{hp}}\kappa_{hp}^{3}\lambda_{hp,t}^{IPFR} \quad (71)$$

$$\rho_{i,t}^{VPP,IN}=\lambda_{n,t}^{IN}+\sum_{hp\in N_{hp}}\kappa_{hp}^{1}\lambda_{hp,t}^{IPFR} \quad (72)$$

$$\rho_{t}^{VPP,DR}=\lambda_{t}^{DR}+\sum_{hp\in N_{hp}}\kappa_{hp}^{4}\lambda_{hp,t}^{IPFR} \quad (73)$$

## VI. Case Studies

In this section, the performance of the proposed VPP aggregation model and the joint energy-reserve-IPFR market mechanism are verified. The VPP frequency response aggregation model is built and validated in MATLAB/Simulink, while the market-clearing problem is solved using the Gurobi solver within a Python environment.

### *A. Basic Data*

The proposed framework is validated on a modified IEEE 30-bus system, designed to represent the high renewable penetration in modern power systems. In this model, the capacities of the SG, wind power, photovoltaic (PV), ES, and VPP are 260 MW, 110 MW, 320 MW, 50 MW, and 220 MW, respectively. The load profile is generated based on historical CAISO data, and a disturbance equivalent to 8% of the total load $P_{\text{load}}$ is applied. The efficiency and initial state-of-charge of the ES units are set to 0.95 and 0.5, respectively. The delays for inertia provision of GFM IBRs ($\tau_1$) and PFR provision of SGs ($\tau_2$) are set to 0.05s and 1s, respectively. The forecast errors for REG $j$ are assumed to follow a normal distribution: $\boldsymbol{\varepsilon}_{i,t}$ ~N($0.05\times P_{j,t}^{REG,EN}$, $0.01\times P_{j,t}^{REG,EN}$), and the confidence level for CCs is set to 95%. Three VPPs are incorporated in the test system, in which the capacities of the affiliated DERs are summarized in Table III. Table IV lists the cost parameters for different services, obtained from [17] and [43]. Notably, the power costs of the VPP are split between small SGs and converter-interfaced DERs to reflect their heterogeneous technologies.

TABLE III
Total Capacities of Different DER Types Aggregated by VPPs

| VPP Number | Small SG (MW) | REG (MW) | ES (MW) | EV (MW) | FL (MW) |
|---|---|---|---|---|---|
| VPP #1 | 30 | 25 | 10 | 5 | 10 |
| VPP #2 | 26 | 10 | 8 | 10 | 8 |
| VPP #3 | 32 | 0 | 23 | 5 | 10 |
| Total | 88 | 35 | 41 | 20 | 28 |

TABLE IV
The Costs for Different Services of Various Providers

| Service Provider | Power ($/MWh) SG | Power ($/MWh) Others | Virtual Inertia ($/(MW·s/Hz)) | Droop Factor ($/(MW/Hz)) |
|---|---|---|---|---|
| SG | 30-35 | - | - | - |
| REG | - | 4-6 | 2-4 | 3-5 |
| ES | - | 12-16 | 2-4 | 3-5 |
| VPP | 32-36 | 10-15 | 2-4 | 3-5 |

### *B. The Aggregation Model of VPP*

The performance of the proposed VPP aggregation model is evaluated using first-, second-, and third-order equivalent transfer functions, with parameters identified by the SGD-based procedure as shown in Table V.

TABLE V
Equivalent Parameters of the VPP Aggregation Models

| Parameters VPP Number | Non-delayed Inertia (MW·s/Hz) | Delayed Inertia (MW·s/Hz) | Droop Factor (MW/Hz) |
|---|---|---|---|
| VPP #1 | 4.1 | 3.3 | 19.5 |
| VPP #2 | 3.8 | 3.0 | 16.8 |
| VPP #3 | 4.3 | 3.1 | 17.5 |

In Fig. 7, the system frequency dynamics produced by the full-order model of VPP #1 are compared with the frequency dynamics produced by its first-, second-, and third-order aggregated equivalents, when a disturbance N($0.08\times P_{\text{load}}$, $0.12\times P_{\text{load}}$) MW occurs at $t$=$0^{+}$. The second- and third-order models track the nadir and QSS frequency of the full-order model within the 95 % confidence interval (CI), while the first-order model is marginally conservative, and yet its maximum absolute error is still below 0.0003 p.u. (50 Hz base).

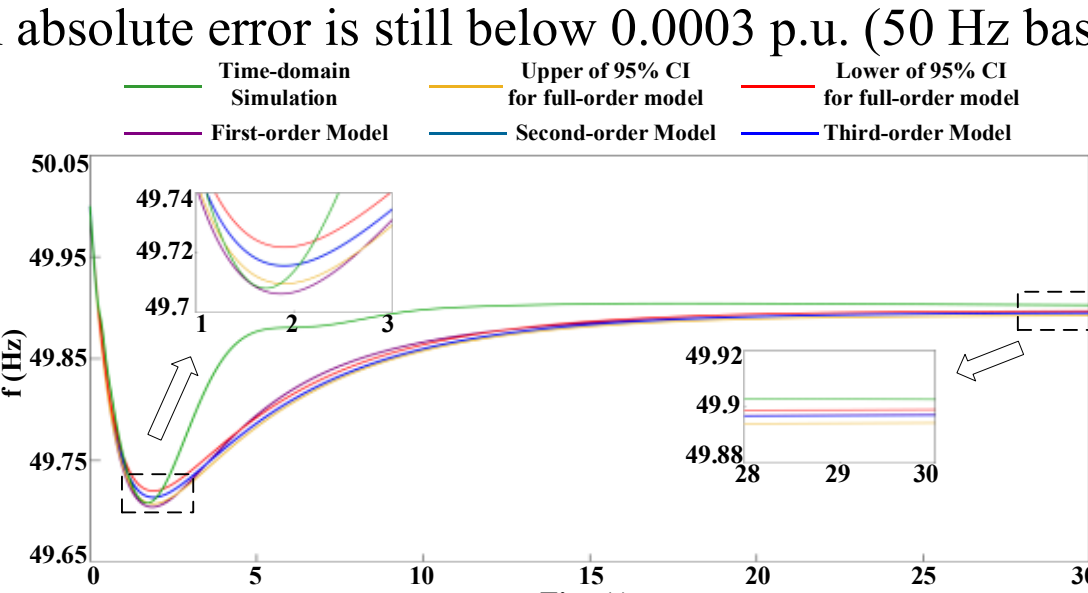

Fig. 7. System frequency dynamics from time-domain simulation and aggregation models for VPP #1.

To further validate the model under electrically-coupled grid conditions, a time-domain simulation is conducted in MATLAB/Simulink using the actual network topology, with DERs placed at their physical locations. The time-domain simulation result (green trace in Fig. 7) demonstrates that the proposed aggregation model matches closely with the electrically-coupled benchmark at the nadir point, with a minor deviation of 0.001 p.u. observed at the QSS point. Thus, the proposed aggregation model introduces negligible errors, even when detailed electrical coupling is considered.

To quantitatively evaluate the impact of inertial response delay, 500 disturbance scenarios are generated by sampling from a normal distribution N($0.08\times P_{\text{load}}$, $0.12\times P_{\text{load}}$). The MAPEs of both $f_{\text{nadir}}$ and $f_{\text{qss}}$ between the proposed aggregation model and the electrically-coupled benchmark are presented in Table VI. The models accounting for delays achieve higher accuracy in capturing the dynamics at the nadir and QSS frequency points.

TABLE VI
MAPEs OF NADIR FREQUENCY AND QSS FREQUENCY USING MODELS WITH AND WITHOUT CONSIDERING DELAY

| Model | Considering Delay | | | Without Considering Delay | | |
|---|---|---|---|---|---|---|
| Indicator | 1-order | 2-order | 3-order | 1-order | 2-order | 3-order |
| MAPE of $f_{\text{nadir}}$ | 0.008% | 0.024% | 0.023% | 0.012% | 0.026% | 0.026% |
| MAPE of $f_{\text{qss}}$ | 0.021% | 0.018% | 0.018% | 0.024% | 0.019% | 0.018% |

A sensitivity analysis is further conducted to evaluate the influence of inertial response delay on the accuracy of the proposed aggregation model. The values of $\tau_1$ are varied within the ranges [0.01, 0.05] s, reflecting typical values reported in [38]. The resulting MAPEs for both $f_{\text{nadir}}$ and $f_{\text{qss}}$ under different delay values are summarized in Table VII. The results demonstrate that the aggregation model maintains accuracy across the considered delay ranges, with MAPEs consistently below 0.03%. This confirms that the proposed method is robust against variations in inertial response delay, enhancing its practical applicability in systems with diverse dynamic characteristics. Besides, the computation times for the first-, second-, and third-order models are 87, 61, and 37 seconds, respectively, thus satisfying the practical requirements for market operations.

TABLE VII
MAXIMUM MAPEs OF DIFFERENT AGGREGATION MODELS UNDER DIVERSE VALUES OF INERTIAL RESPONSE DELAY

| Value of $\tau_1$ | 0.01s | 0.02s | 0.03s | 0.04s | 0.05s |
|---|---|---|---|---|---|
| MAPE of $f_{\text{nadir}}$ | 0.026% | 0.024% | 0.024% | 0.024% | 0.028% |
| MAPE of $f_{\text{qss}}$ | 0.022% | 0.021% | 0.022% | 0.020% | 0.020% |

## C. *Market-clearing Results*

The clearing results for the joint energy-reserve-IPFR market are presented in Figs. 8-11. During periods of low REG output power (0–6 h and 19–23 h), VPPs offer energy at a lower marginal cost than SGs and are consequently dispatched (Fig. 8(a)). Conversely, REGs are prioritized when their output power is sufficient (7–18 h). Consequently, energy prices are higher in hours 0–6 and 19–23, when PV generation is low and VPPs set the marginal price (Fig. 8(b)).

In the reserve market, VPPs and ES units are the primary contributors (Fig. 9(a)). VPPs notably provide substantial reserve capacity during periods of high REGs' generation. As a result of this abundant and flexible capacity, the clearing price remains flat throughout most operating hours (Fig. 9(b)).

Inertia is cleared only by SGs and VPPs (Fig. 10(a)). The clearing price (Fig. 10(b)) for inertia exhibits a significant spike during periods of inertia shortage, which usually occur during the periods of high REG output power, i.e., 7–18 h.

ES units and VPPs account for a considerable share of droop factor provision (Fig. 11(a)). ES units are highly cost-effective at supplying droop factors due to the decoupling between high-cost energy and low-cost droop services. VPPs also provide substantial droop support, leveraging their low cost and high dispatchability. The droop factor price remains stable except for a notable surge between 19–21 h (Fig. 11(b)). This is because a simultaneous shortage in system droop capacity forces the activation of high-cost marginal SGs.

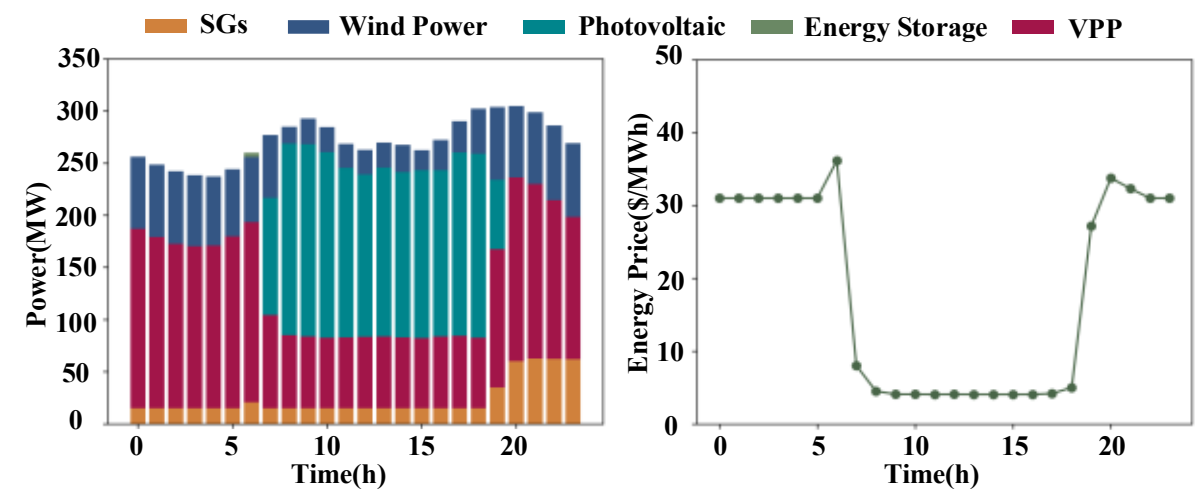

(a) Power mix (b) Clearing price for energy

Fig. 8. Power provision and clearing price for the IEEE 30-bus system.

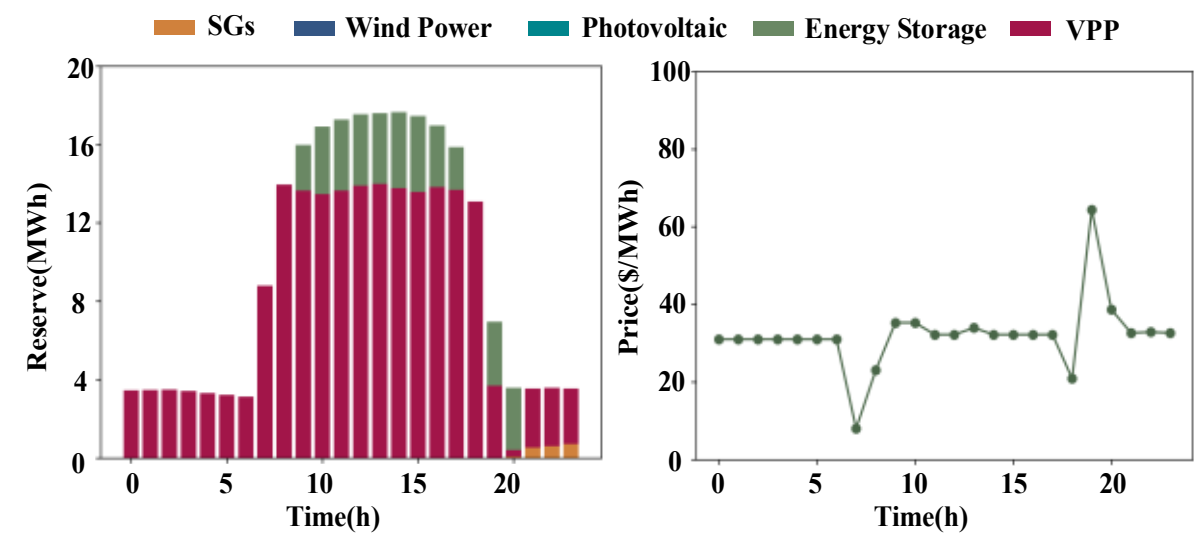

(a) Reserve mix (b) Clearing price for reserve

Fig. 9. Reserve provision and clearing price for the IEEE 30-bus system.

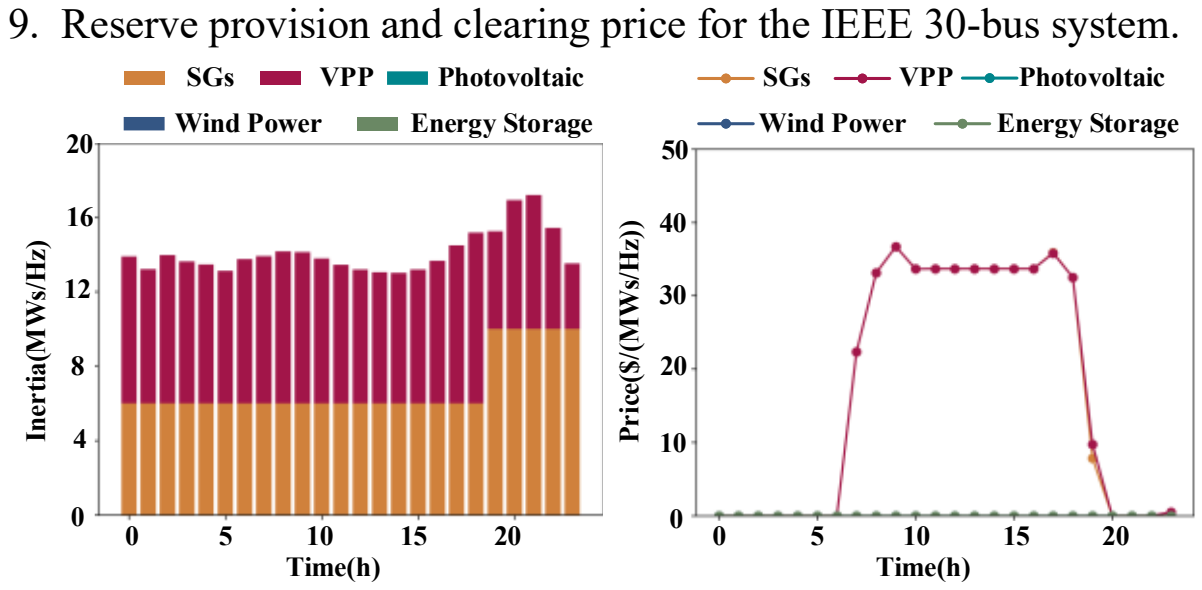

(a) System inertia mix (b) Clearing price for inertia

Fig. 10. Inertia provision and clearing price for the IEEE 30-bus system.

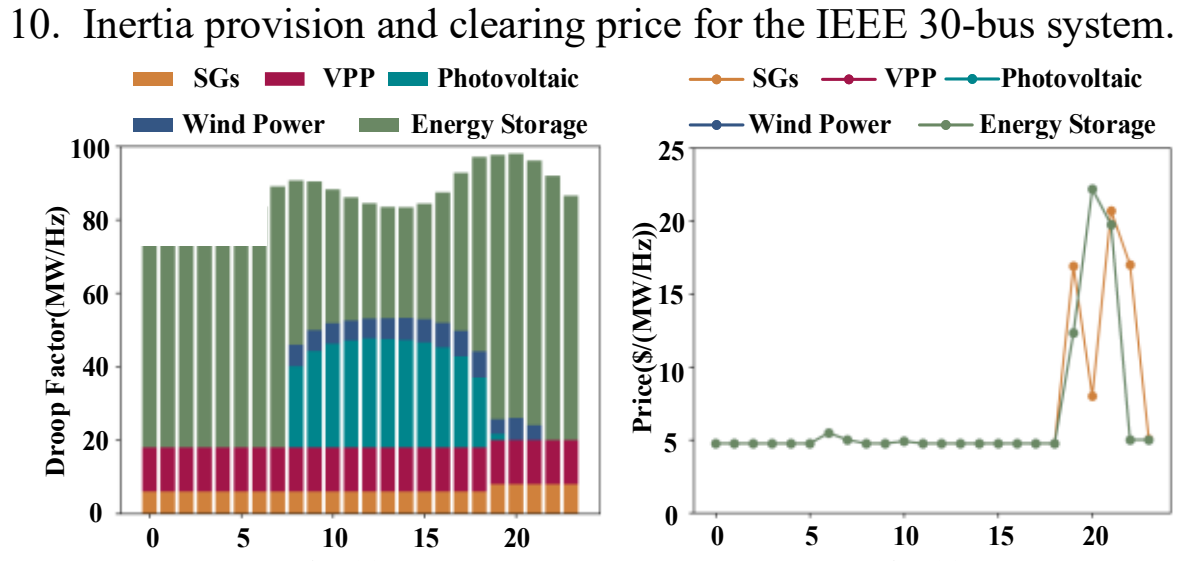

(a) System droop factor mix (b) Clearing price for droop factor

Fig. 11. Droop factor provision and clearing price for the IEEE 30-bus system.

## D. *Benefits of VPP Participation*

To quantify the benefits of participation of VPPs, two scenarios are compared: Scenario 1 (basic case with VPPs) and Scenario 2 (VPPs deactivated, with other resources scaled up proportionally). A direct comparison of the market-clearing results (Figs. 9–10 and 12–13) shows that in the absence of VPPs, SGs are required to supply significantly more reserve and inertia, leading to higher reserve and inertia prices during periods of high REGs' generation. In contrast, the participa-tion of VPPs provides a substantial share of lower-cost reserve and non-delayed inertia, which reduces total system cost and stabilizes market prices for reserve and inertia.

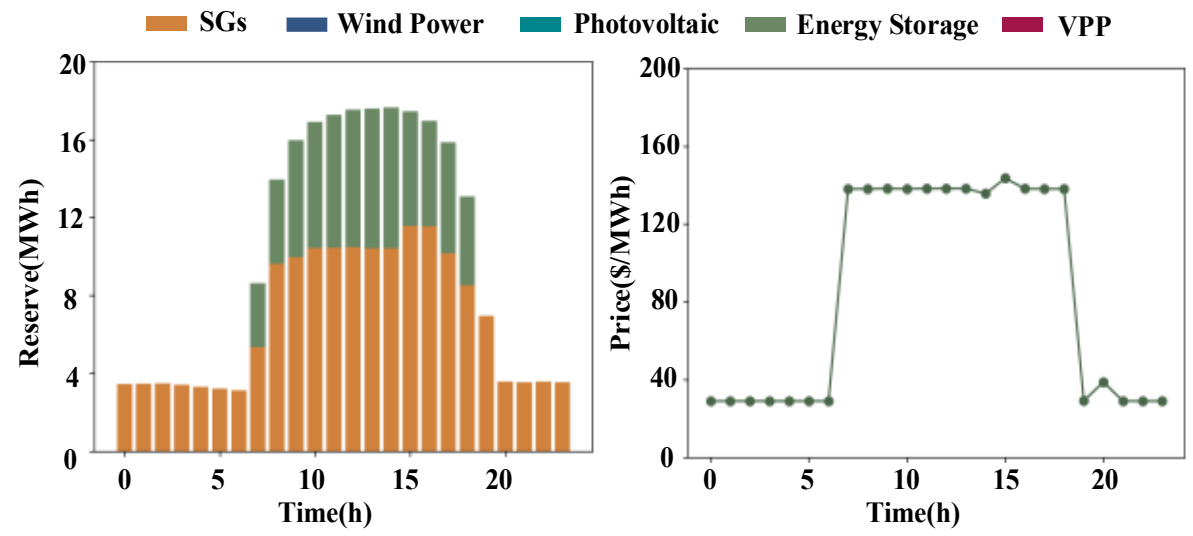

(a) Reserve mix (b) Clearing price for reserve

Fig. 12. Reserve provision and clearing price for Scenario 2.

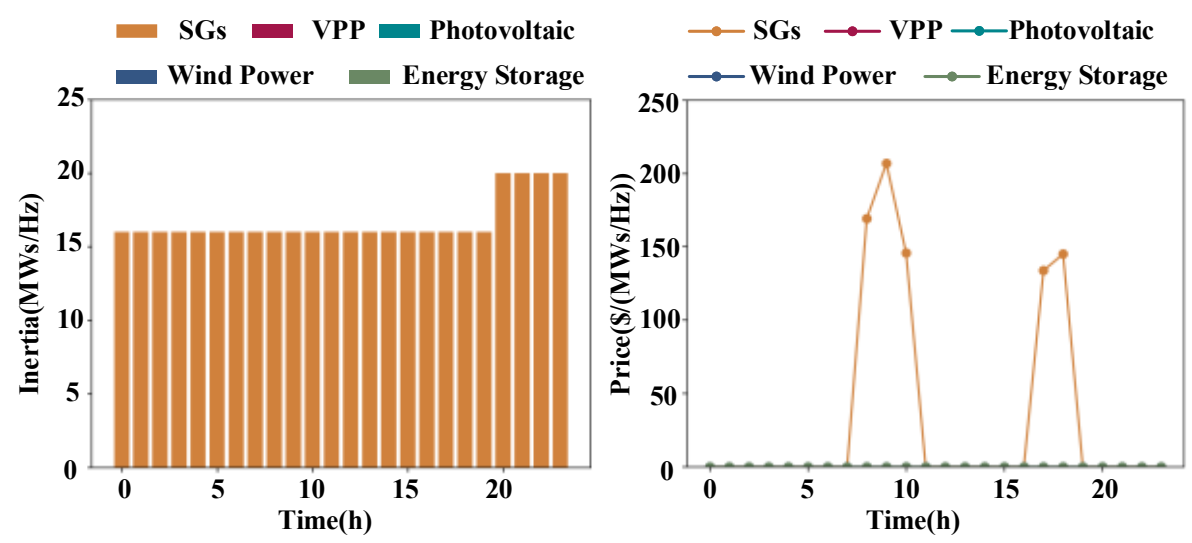

(a) System inertia mix (b) Clearing price for inertia

Fig. 13. Inertia provision and clearing price for Scenario 2.

The costs and net profits are summarized in Table VIII. The comparison demonstrates that the participation of VPPs enhances system flexibility in energy and inertia provision, reducing total costs by approximately 40% and increasing net profit by about 30%. Furthermore, the costs of SGs are substantially reduced because VPPs diminish the need for frequent changes of SG status to address inertia shortages and reduce dependence on SG power outputs.

TABLE VIII
COST AND NET PROFIT OF THE JOINT ENERGY-RESERVE-IPFR MARKET WITH AND WITHOUT VPPS FOR THE IEEE 30-BUS SYSTEM

| Service Provider | Scenario 1 | | Scenario 2 | |
|---|---|---|---|---|
| | Cost($) | Net Profit($) | Cost($) | Net Profit($) |
| SG | 19,601 | -2,769 | 124,488 | -10,813 |
| VPP | 41,593 | 33,477 | - | - |
| REG | 14,456 | 25,505 | 12,589 | 30,648 |
| ES | 7,044 | 3,412 | 9,303 | 25,038 |
| Total | 82,696 | 59,625 | 146,381 | 44,873 |

### *E. Scalability to Larger Systems*

The scalability and practical applicability of the proposed framework are further validated on a modified IEEE 118-bus system [12], with a 35% REG penetration representing a high-renewable scenario. The test system is integrated with 13 VPPs, each possessing diverse DER portfolios, including 100% renewable VPPs and hybrid VPPs, to compare their market participation [44]. The stochastic setting follows the 30-bus case, with more detailed parameters for the IEEE 118-bus system available in [45].

The proposed VPP aggregation model is applied to each of the 13 VPPs, whose aggregated capacities range from 85 MW to 477 MW. Market-clearing is then performed using the Gurobi solver. The total computational time for both the aggregation process and market-clearing remains under 3 minutes, demonstrating the computational efficiency of the approach for large-scale systems.

To assess the accuracy of the aggregation models across the diverse VPP portfolio, 500 scenarios with different disturbance values are simulated to obtain the frequency dynamics of the system. The absolute values of the nadir and QSS frequency errors for the first-, second-, and third-order aggregation models of each VPP are plotted in Fig. 14. The results indicate that the proposed aggregation model maintains acceptable accuracy across the wide range of VPP capacities. Notably, the 100% renewable VPPs (located at buses 27, 56, 62, and 91), by virtue of comprising fewer types of DERs and thus exhibiting lower-order dynamics, demonstrate relatively smaller aggregation errors than the hybrid VPPs.

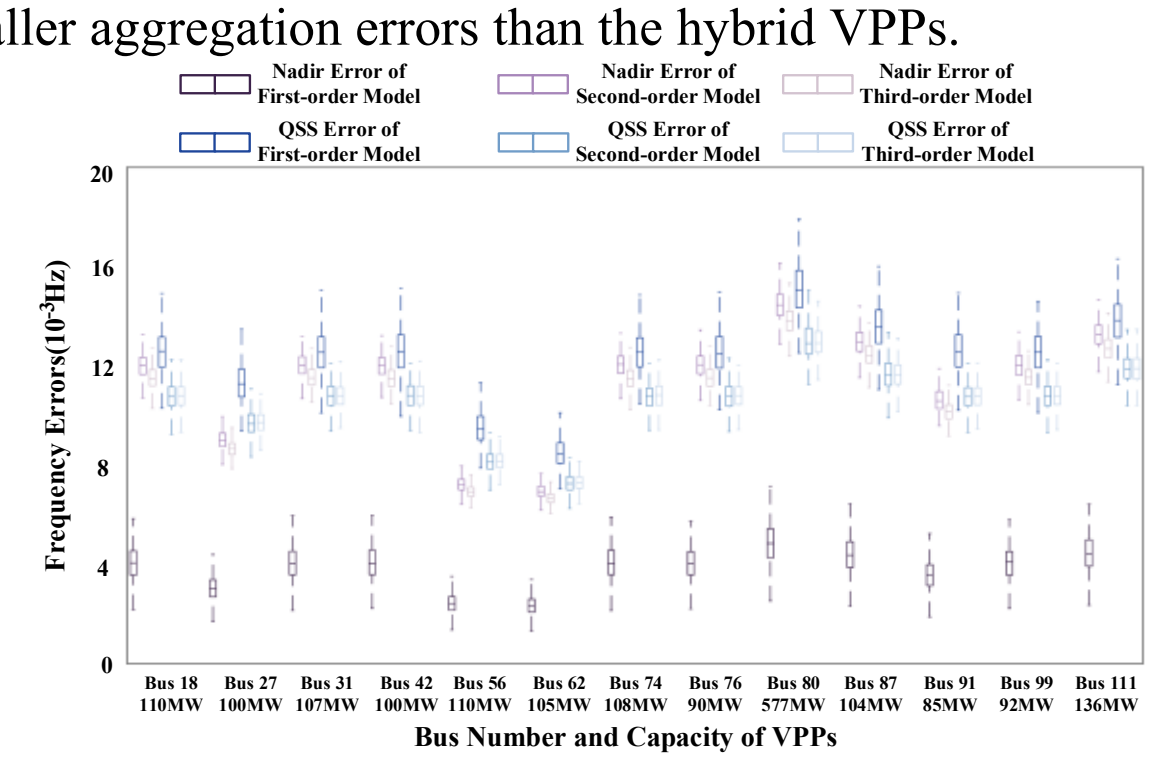

Fig. 14. Absolute values of nadir and QSS frequency errors of each VPP.

Market-clearing results for the IEEE 118-bus system are illustrated in Figs. 15-18. For comparison, Scenario 3 (base case with VPPs) and Scenario 4 (VPPs deactivated, with other resources scaled up proportionally) are defined, with their costs and net profits compared in Table IX. These results indicate that VPPs continue to provide cost-effective inertia and droop factors, while the prices for inertia and droop factors exhibit similar trends to the 30-bus case, confirming the stability and scalability of the proposed pricing mechanism.

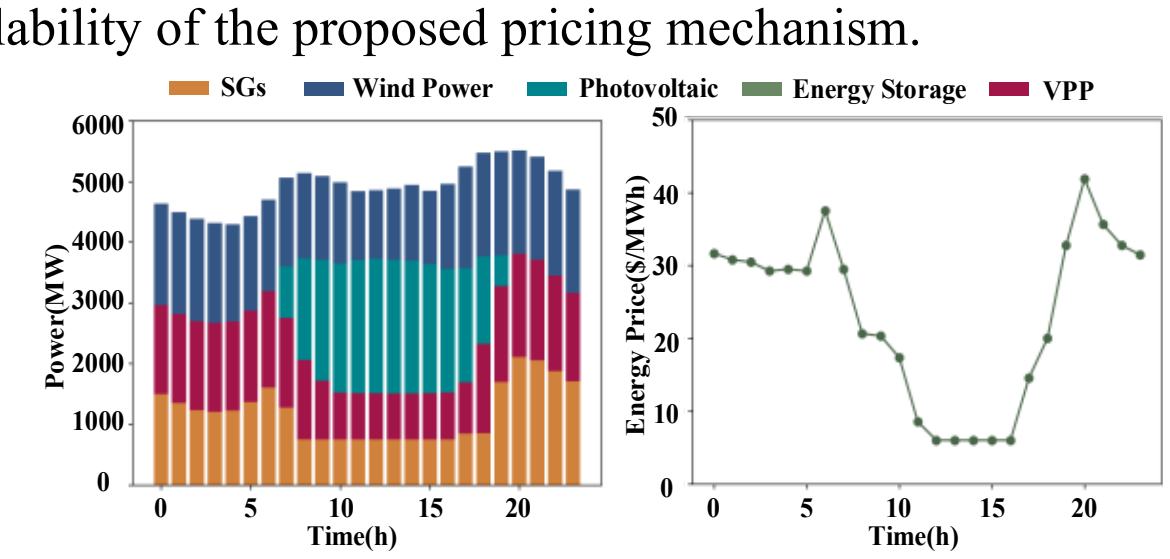

(a) Power mix (b) Clearing price for energy

Fig. 15. Power provision and clearing price for the IEEE 118-bus system.

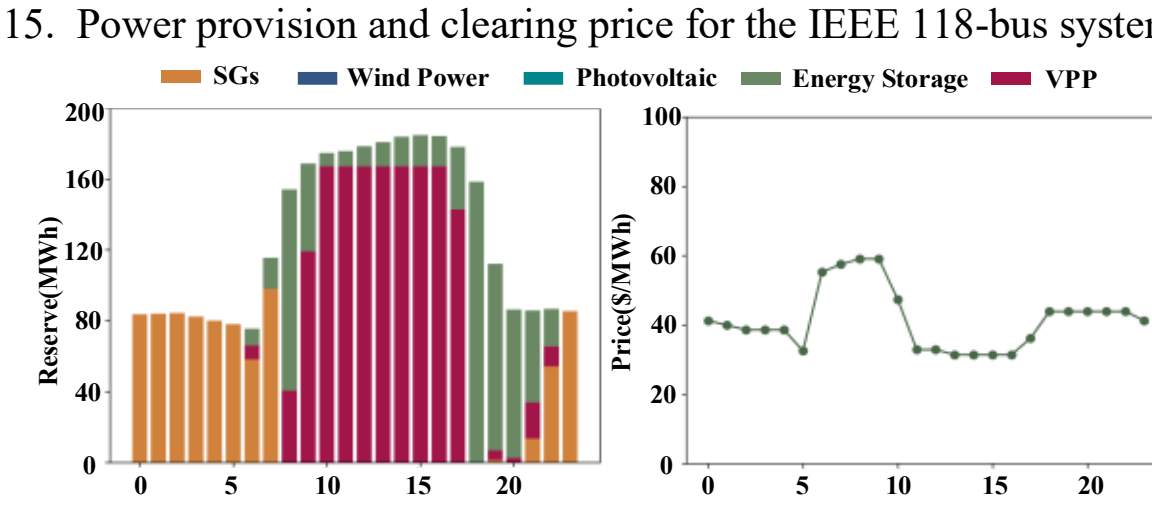

(a) Reserve mix (b) Clearing price for reserve

Fig. 16. Reserve provision and clearing price for the IEEE 118-bus system.

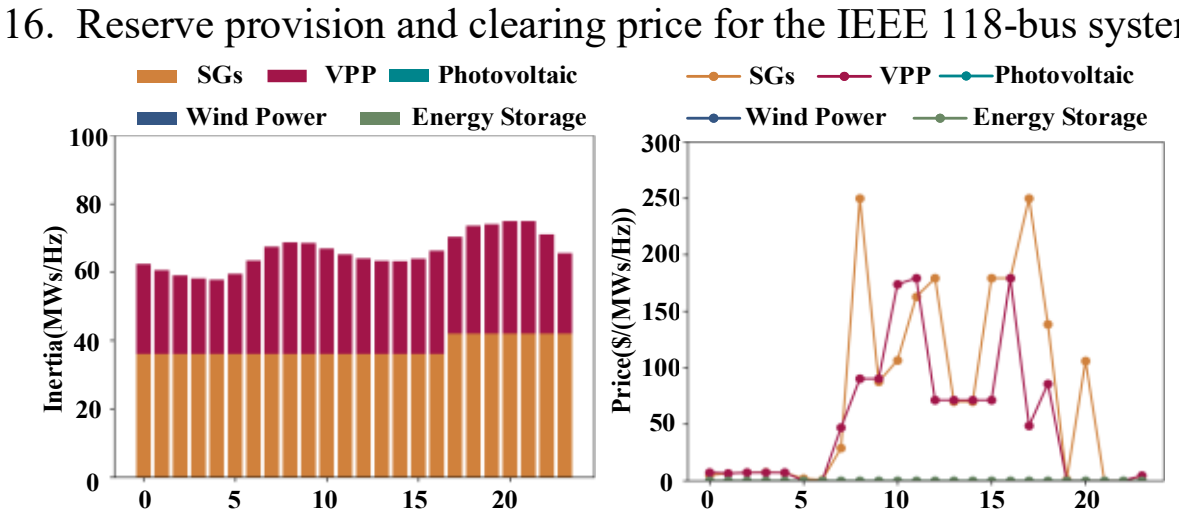

(a) System inertia mix (b) Clearing price for inertia

Fig. 17. Inertia provision and clearing price for the IEEE 118-bus system.

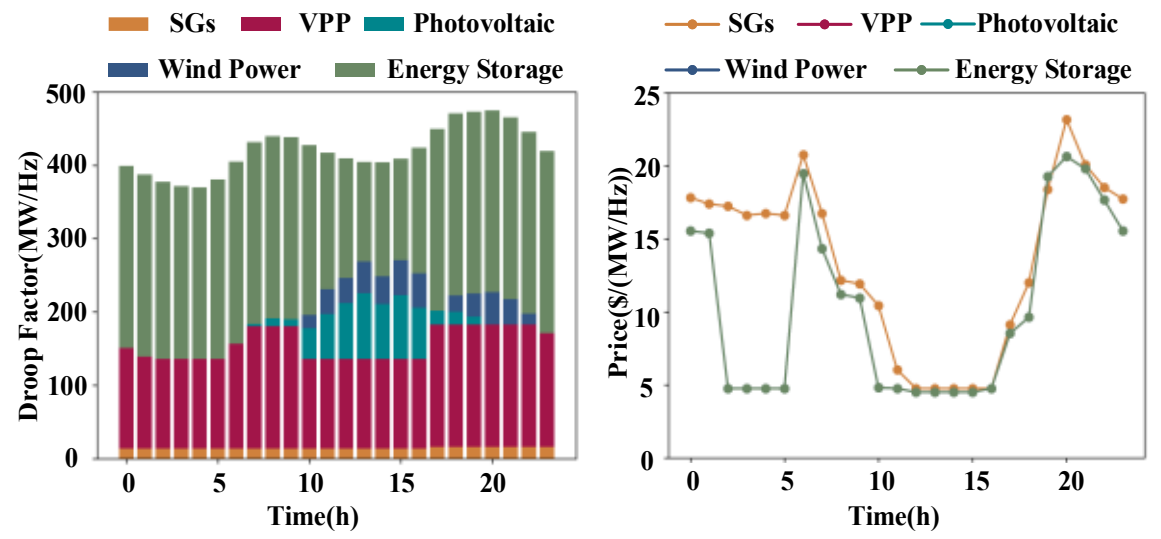

(a) System droop factor mix (b) Clearing price for droop factor

Fig. 18. Droop factor provision and clearing price for the IEEE 118-bus system.

TABLE IX
COST AND NET PROFIT OF THE JOINT ENERGY-RESERVE-IPFR MARKET WITH AND WITHOUT VPPS FOR THE IEEE 118-BUS SYSTEM

| Service Provider | Scenario 3 | | Scenario 4 | |
|---|---|---|---|---|
| | Cost($) | Net Profit($) | Cost($) | Net Profit($) |
| SG | 866,452 | 19,178 | 192,045 | 16,358 |
| VPP | 353,007 | 557,218 | - | - |
| REG | 279,486 | 901,132 | 286,762 | 1,068,409 |
| ES | 45,382 | 47,742 | 80,369 | 102,200 |
| Total | 1,544,328 | 1,525,272 | 2,152,596 | 1,186,968 |

The services portfolio for each VPP in the joint energy-reserve-IPFR market is shown in Fig. 19. It reveals that 100% renewable VPPs are excluded from the inertia and reserve markets due to delays in virtual inertia provision and the stochastic nature of REGs. Nevertheless, they achieve profitability through participation in the energy and PFR markets. This economic distinction is quantified in Table X, which compares the costs and benefits of a 100% renewable VPP (bus 27) and a hybrid VPP (bus 42) with the same capacity. The result reveals that the 100% renewable VPP maintains a cost and profit advantage in the energy market.

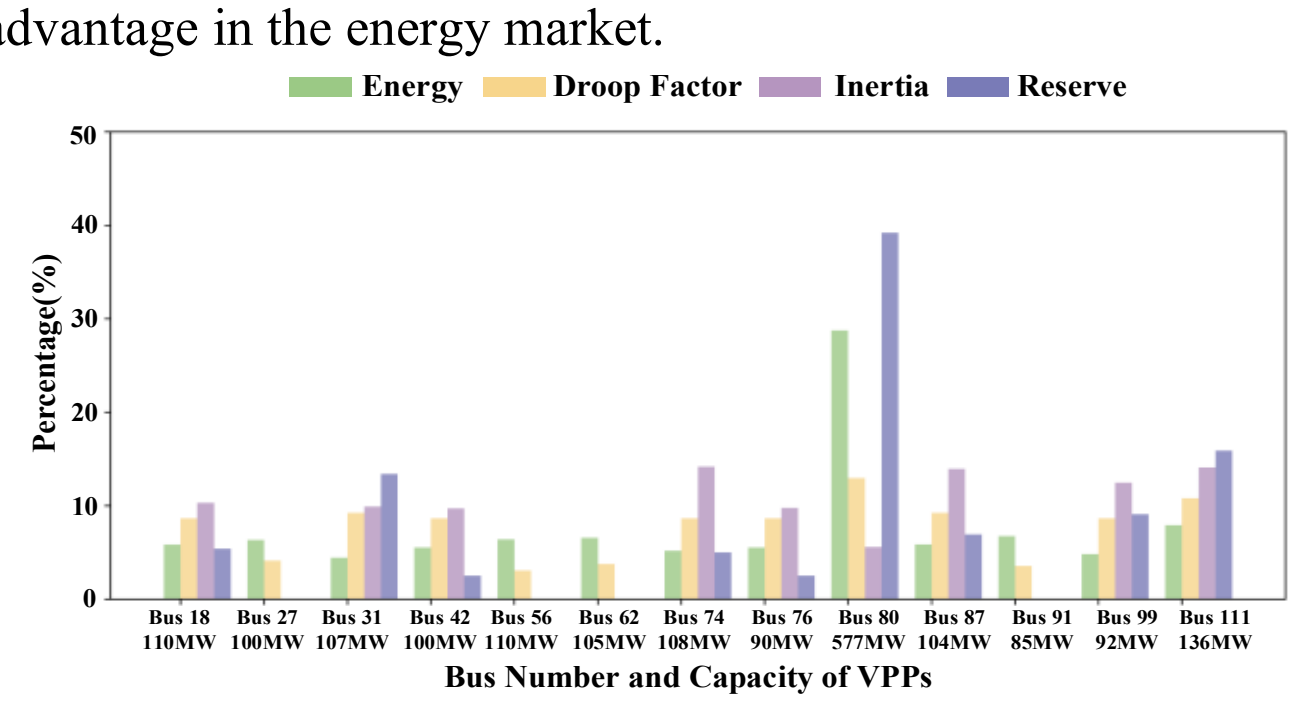

Fig. 19. The percentage of the total value of services provided by each VPP.

TABLE X
THE COSTS AND BENEFITS OF 100% RENEWABLE VPP AT BUS 27 AND HYBRID VPP AT BUS 42

| Bus | Energy | | Droop Factor | | Inertia | | Reserve | |
|---|---|---|---|---|---|---|---|---|
| | Cost($) | Profit($) | Cost($) | Profit($) | Cost($) | Profit($) | Cost($) | Profit($) |
| 27 | 14,662 | 47,790 | 621 | 2,263 | - | - | - | - |
| 42 | 16,370 | 43,091 | 1,152 | 2,991 | 0 | 3,654 | 3,185 | 3,758 |

These numerical results on the large-scale test system demonstrate that the proposed framework remains effective and scalable for complex power systems. Performance and computational efficiency are maintained under conditions of high renewable penetration and highly heterogeneous resource portfolios.

## VII. CONCLUSION

A novel VPP aggregation model and a joint energy-reserve-IPFR market mechanism are proposed in this paper to address frequency stability challenges in low-inertia power systems. An optimization-based model that aggregates heterogeneous DERs is established, accounting for inertial response delays. The energy-reserve-IPFR market mechanism is designed considering REGs' generation uncertainty to facilitate the participation of diverse resources in providing the IPFR product. Numerical results validate the effectiveness of the VPP aggregation model and demonstrate that the energy-reserve-IPFR market mechanism successfully facilitates DERs' participation, fully leveraging the advantages of VPPs in providing schedulable, non-delayed inertia.

In future work, the VPP aggregation model will be extended to incorporate internal device limits, thereby enabling a detailed investigation into the strategic bidding of aggregated DER portfolios and its associated incentives and settlement implications.

## APPENDIX

### A. The Derivation of the System Frequency Formula

This appendix details the derivation of the piecewise function with 3 cases for system frequency deviation, following a disturbance at $t$=0$^+$.

*1) Case 1: Initial Inertia Response ($t = 0^+$)*

The system frequency dynamics for Case 1 are governed by the swing equation. The frequency deviation $\Delta f_1(t)$ is described by the solution to the differential equation obtained by substituting (1) and (13) into (15)-(16), yielding (A1). By applying the initial condition $\Delta f_1(0)$ =0 into (A1), the frequency deviation $\Delta f_1(t)$ is obtained as (A2).

$$2H^{GV}\cdot\frac{d\Delta f_1(t)}{dt}=-\Delta D, 0<t\le\tau_1 \tag{A1}$$

$$\Delta f_1(t)=\frac{\Delta D}{2H^{GV}}t,\quad t=0^+ \tag{A2}$$

*2) Case 2: Activation of Delayed Resources ($0 < t \le \tau_2$)*

For $t > 0^+$, GFM IBRs and VPPs begin to inject power $\Delta P_\tau$ into the system. As justified in Section IV-B, the delay $\tau_1$ associated with GFM IBRs is negligible for this derivation. Consequently, by substituting (11)-(12) into (15)-(16), (A3) is obtained, with the general solution given by (A4). By applying the initial condition $\Delta f_2(0)$ =0, obtains $\Delta f_2(t)$ as (A5).

$$\begin{cases}2H^{TO}\dfrac{d\Delta f_1(t)}{dt}+\left(k^{FM}+k^{VPPR}\right)\Delta f_1(t)=-\Delta D\,,\ 0<t\le\tau_2\\ k^{FM}=\sum\limits_{k\in\Omega_{ESS}}k_k^{ESS}+\sum\limits_{j\in\Omega_{REG}}k_j^{REG}\\ k^{VPPR}=\sum\limits_{l\in\Omega_{VPP}}k_j^{VPPR}\end{cases} \tag{A3}$$

$$\begin{aligned}\Delta f_2(t)&=e^{-\int\frac{k^{FM}+k^{VPPR}}{2H^{TO}}dt}\left(C+\int-\frac{\Delta D}{2H^{TO}}e^{\int\frac{k^{FM}+k^{VPPR}}{2H^{TO}}dt}dt\right)\\&=-\frac{\Delta D}{k^{FM}+k^{VPPR}}+Ce^{-\frac{k^{FM}+k^{VPPR}}{2H^{TO}}t},0<t\le\tau_2\end{aligned} \tag{A4}$$

$$\Delta f_2(t)=\frac{\Delta D}{k^{FM}+k^{VPPR}}\left(e^{-\frac{k^{FM}+k^{VPPR}}{2H^{TO}}t}-1\right),0<t\le\tau_2 \tag{A5}$$

During 0< $t \le\tau_2$, the GFM IBRs, and VPPs have already provided a portion of the power $\Delta P_\tau$ to the system. At $t = \tau_2$, a power deficiency $\Delta D'<\Delta D$ is present. The calculations for $\Delta P_\tau$ and $\Delta D'$ are derived in (A6)-(A7).

$$\begin{aligned}\Delta P_\tau &= -2\left(H^{FM}+H^{VPPR}\right)\Delta f_2'(t)-\left(k^{FM}+k^{VPPR}\right)\Delta f_2(t)\Big|_{t=\tau_2}\\ &= \Delta D\left(1-\frac{H^{GV}}{H^{TO}}\cdot e^{-\frac{k^{FM}+k^{VPPR}}{2H^{TO}}\tau_2}\right)\end{aligned} \tag{A6}$$

$$\Delta D' = \Delta D - \Delta P_\tau = \frac{H^{GV}}{H^{TO}}\cdot e^{-\frac{k^{FM}+k^{VPPR}}{2H^{TO}}\tau_2}\Delta D \tag{A7}$$

*3) Case 3: Full Primary Frequency Response ($t > \tau_2$)*

For $t > \tau_2$, the PFR from SGs is fully active. Substituting (1) and (11)-(13) into (15)-(16) yields a second-order differential equation as shown below:

$$\begin{aligned}-\Delta D' &= 2H^{TO}T^{GV}\cdot\frac{d^2\Delta f_3(t)}{dt^2}+\left(k^G+k^{FM}+k^{VPP}\right)\Delta f_3(t)\\ &+\left(2H^{GV}+k^{FM}T^{GV}+k^{VPP}T^{GV}\right)\cdot\frac{d\Delta f_3(t)}{dt}\end{aligned} \tag{A8}$$

$$\begin{cases}T^{GV}=\sum_{i\in\Omega_G}\lambda_i T_i^G+\sum_{l\in\Omega_{VPP}}\lambda_l T_l^{VPP}\\ k^G=\sum_{i\in\Omega_G}x_{i,t}k_i^G,\ k^{VPP}=k^{VPPR}+\sum_{l\in\Omega_{VPP}}x_{l,t}k_l^{VPPG}\end{cases} \tag{A9}$$

Solving the differential equation (A8), the characteristic roots, special solution, and general solution are obtained as (A10)-(A12). Substituting the general solution into (A13), yields (A14).

$$\begin{aligned}r_{1,2} &= -\frac{2H^{GV}+\left(k^{FM}+k^{VPP}\right)T^{GV}}{4H^{TO}T^{GV}}\\ &\pm\sqrt{\left(\frac{2H^{GV}+k^{FM}T^{GV}+k^{VPP}T^{GV}}{4H^{TO}T^{GV}}\right)^2-\frac{k^G+k^{FM}+k^{VPP}}{2H^{TO}T^{GV}}}\end{aligned} \tag{A10}$$

$$\Delta f_3^*(t) = -\frac{\Delta D'}{k^G+k^{FM}+k^{VPP}} \tag{A11}$$

$$\begin{cases}\Delta f_3(t)=\Delta D' e^{-\alpha t}[C_1\sin(\omega t)+C_2\cos(\omega t)]-\dfrac{\Delta D'}{k^G+k^{FM}+k^{VPP}}\\ \alpha=\dfrac{2H^{TO}+\left(k^{FM}+k^{VPP}\right)T^{GV}}{4T^{GV}H^{TO}}\\ \omega=\sqrt{\left|\dfrac{k^G+k^{FM}+k^{VPP}}{2T^{GV}H^{TO}}-\alpha^2\right|}\end{cases} \tag{A12}$$

$$\begin{cases}\Delta f_2(t)\big|_{t=\tau_2}=\Delta f_3(t)\big|_{t=\tau_2^+}\\ \dfrac{d\Delta f_2(t)}{dt}\Big|_{t=\tau_2}=\dfrac{d\Delta f_3(t)}{dt}\Big|_{t=\tau_2^+}\end{cases} \tag{A13}$$

$$\begin{cases}\dfrac{\Delta D}{k^{FM}+k^{VPP}}\left(e^{-\frac{k^{FM}+k^{VPP}}{2H^{TO}}\tau_2}-1\right)=-\dfrac{\Delta D'}{k^G+k^{FM}+k^{VPP}}\\ \qquad +\Delta D' e^{-\alpha\tau_2}[C_1\sin(\omega\tau_2)+C_2\cos(\omega\tau_2)]\\ -\dfrac{\Delta D}{2H^{TO}}e^{-\frac{k^{FM}+k^{VPP}}{2H^{TO}}\tau_2}=\Delta D' e^{-\alpha\tau_2}[C_1\omega\cos(\omega\tau_2)-C_2\omega\sin(\omega\tau_2)]\\ \qquad -\Delta D'\alpha e^{-\alpha\tau_2}[C_1\sin(\omega\tau_2)+C_2\cos(\omega\tau_2)]\end{cases} \tag{A14}$$

Since the $\Delta f_2'(t) < 0$ for $t \le \tau_2$, $\Delta f_2(t)$ decreases monotonically, and thereby the frequency nadir occurs during $t > \tau_2$.

## BIOGRAPHIES

**Changsen Feng** (Member, IEEE) received the B.E. degree in electrical engineering from Shandong University, Jinan, China, in 2013, and the Ph.D. degree from Zhejiang University, Hangzhou, China, in 2019. He joined the Faculty of the Zhejiang University of Technology in July 2019 and is currently an Assistant Professor with the College of Information Engineering. His research interests include game theory, machine learning, and optimization theory in power systems.

**Zhongliang Huang** received the B.E. degree in electrical engineering from the Zhejiang University of Technology, Hangzhou, China, in 2023. He is currently working toward the M.E. degree in control science and engineering with the Zhejiang University of Technology. His research interests include virtual power plants, power system economics and optimization.

**Jun Lin** received the B.E. degree in electrical engineering from the Quzhou University, Quzhou, China, in 2022, and the M.E. degree in control science and engineering from the Zhejiang University of Technology in 2025. His research interests include electricity markets.

**Licheng Wang** received the B. Eng. degree from the Huazhong University of Science and Technology, Wuhan, China, in 2012, the M. Eng. degree from Zhejiang University, Hangzhou, China, in 2015, and the Ph.D. degree from the University of Queensland, Brisbane, QLD, Australia, in 2019. He is currently a tenured Associate Professor with the College of Information Engineering, Zhejiang University of Technology, Hangzhou. His research interests include power system operation and control, renewable energy integration into distribution systems, distributed algorithms, and deep reinforcement learning and its application in networked systems.

**Youbing Zhang** (Senior Member, IEEE) received the B.S. and M.S. degrees in electrical engineering from Hunan University, Changsha, China, in 1993 and 1996, respectively, and the Ph.D. degree in electrical engineering from Huazhong University of Science and Technology, Wuhan, China, in 2003. From 1996 to 1997, he was a research assistant both at Tsinghua University, Beijing, China, and China Electric Power Research Institute, Beijing, China. From 1997 to 1999, he was a lecturer at the Department of Electrical Engineering in Hunan University. From 2003 to 2005, he was a postdoctoral research fellow at the Zhejiang University, Hangzhou, China. He is currently a professor with the College of Information Engineering, Zhejiang University of Technology, Hangzhou, China. His research interests include demand-side management, vehicle-to-grid technology, power quality monitoring and control, smart energy, and big data applications.

**Fushuan Wen** (Fellow, IEEE) received the B.E. and M.E. degrees in electrical engineering from Tianjin University, Tianjin, China, in 1985 and 1988, respectively, and the Ph.D. degree in electrical engineering from Zhejiang University, Hangzhou, China, in 1991. In 1991, he joined Zhejiang University as a Faculty Member with the School of Electrical Engineering, where he has been a Full Professor since 1997; he has also been a Full Professor with the Hainan Institute, Zhejiang University, since 2022. He is currently a Visiting Principal Research Scientist with the Shenzhen Institute of Artificial Intelligence and Robotics for Society, Shenzhen, China. He has been listed in "Most Cited Chinese Researchers" in nine consecutive years since 2015 by Elsevier. His research interests include power industry restructuring, power system alarm processing, fault diagnosis and restoration strategies, smart grids and electric vehicles, and artificial intelligence applications in power and integrated energy systems. He is the Editor-in-Chief of Energy Conversion and Economics (IET), Deputy Editor-in-Chief of Automation of Electric Power Systems, and is the editor, subject editor and associate editor of several international journals.